\documentclass[12pt]{article}
\pdfoutput=1
\usepackage{graphicx}
\usepackage{color}
\usepackage{hyperref}
\usepackage[super,compress]{cite}
\usepackage{amsmath}
\usepackage{amssymb}

\newcommand{\comment}[1]{}
\newcommand{\susy}{{\mbox{\scriptsize{SUSY}}}}
\newcommand{\mssm}{{\mbox{\scriptsize{MSSM}}}}
\newcommand{\ncmssm}{{\mbox{\scriptsize{ncMSSM}}}}
\newcommand{\sm}{{\mbox{\scriptsize{SM}}}}
\newcommand{\lr}{{\mbox{\scriptsize{LR}}}}
\newcommand{\ps}{{\mbox{\scriptsize{PS}}}}

\begin{document}
\title{Bayesian naturalness, simplicity, and testability applied to the $B-L$ MSSM GUT}
\author{Panashe Fundira and Austin Purves\\
\\
\it \small Department of Physics, Manhattanville College\\
\it \small Purchase, New York 10577, United States\\
\\
\small fundirap@gmail.com\quad austin.purves@mville.edu
}
\maketitle

\begin{abstract}
Recent years have seen increased use of Bayesian model comparison to quantify notions such as naturalness, simplicity, and testability, especially in the area of supersymmetric model building. After demonstrating that Bayesian model comparison can resolve a paradox that has been raised in the literature concerning the naturalness of the proton mass, we apply Bayesian model comparison to GUTs, an area to which it has not been applied before. We find that the GUTs are substantially favored over the non-unifying puzzle model. Of the GUTs we consider, the $B-L$ MSSM GUT is the most favored, but the MSSM GUT is almost equally favored.

\end{abstract}
\newpage
\tableofcontents

\section{Introduction}
The naturalness principle --- the principle that a correct theory should not require fine-tuning of its parameters to agree with experimental data --- has been widely used in physics both to make predictions and to guide theoretical study\cite{Giudice:2008bi, Nelson}. When a theory does require fine-tuning to be in agreement with experimental data, it is in conflict with the naturalness principle and is said to have a fine-tuning problem. For example, the horizon problem and the flatness problem are both fine-tuning problems in the big bang theory. Together, they motivated the discovery of the theory of inflation\cite{Guth:1980zm} and the theory of the ekpyrotic universe\cite{Khoury:2001wf}. Notable examples of contemporary fine-tuning problems include the cosmological constant problem\cite{Bousso:2007gp,Martin:2012bt}, the strong CP problem\cite{Peccei:2006as}, and the little hierarchy problem of the Minimal Supersymmetric Standard Model (MSSM)\cite{Martin:1997ns, Barbieri:2000gf}. Each of these examples has inspired many new theoretical developments. See Ref.~\citen{Nobbenhuis:2006yf} for a review of possible solutions inspired by the cosmological constant problem. The strong CP problem has inspired theories of axions\cite{Peccei:2006as,Kim:2008hd,Kim:1979if,Shifman:1979if,Dine:1981rt}, and a variety of other possible solutions\cite{Diaz-Cruz:2016pmm, Banerjee:2000qw, Blinov:2016kte, Hook:2014cda, Swain:2010rr}\footnote{See Refs.~\citen{Diaz-Cruz:2016pmm,Swain:2010rr} for brief lists of possible solutions to the strong CP problem.}. The little hierarchy problem has helped inspire the Super-Little Higgs theory \cite{Csaki:2005fc,Bellazzini:2009ix}, the Extended MSSM \cite{Babu:2008ge}, and the NMSSM \cite{Dermisek:2005ar}. Perhaps the most influential of fine-tuning problems is that of fine-tuning in the mass of the standard model (SM) Higgs boson. This is known as the (big) hierarchy problem \cite{Giudice:2008bi, Nelson, Grinbaum:2009sk}.\footnote{It has been argued that the big hierarchy problem is not a true fine-tuning problem because it is dependent on how quadratic divergences are regularized. For example, see Ref.~\citen{Farina:2013mla}.}

The lack of experimental evidence of new physics from recent particle physics experiments such as the LHC has left us with a plurality of models that are consistent with experimental data. This has lead to increasing reliance upon naturalness both for making predictions, and for motivating various theories of physics beyond the standard model, especially supersymmetry. In order to make quantitative predictions using the naturalness principle, it is necessary to have a quantitative definition of naturalness and a criterion to represent the naturalness principle. This arrived in the form of the Barbieri-Giudice (BG) sensitivity, introduced in Refs.~\citen{Ellis:1986yg,Barbieri:1987fn}. The BG sensitivity is a measure of fine-tuning, so that large BG sensitivity corresponds to a lack of naturalness. That is,
\begin{eqnarray}
\mbox{large BG sensitivity} \quad\Leftrightarrow& \mbox{fine-tuned} &\Leftrightarrow\quad \mbox{unnatural}\ ,\nonumber\\
\mbox{no large BG sensitivity} \quad\Leftrightarrow& \mbox{not fine-tuned} &\Leftrightarrow\quad \mbox{natural}\ .\nonumber
\end{eqnarray}
This became a dominant way to quantify fine-tuning (for example, see Refs.~\citen{Antoniadis:2014eta,Ciafaloni:1996zh,de Carlos:1993yy,Casas:2003jx,Casas:2004gh,Casas:2005ev,Allanach:2006jc,Giusti:1998gz}). However, some shortcomings of the BG sensitivity have been identified, suggesting that it is not appropriate for every situation \cite{Grinbaum:2009sk,de Carlos:1993yy, Anderson:1994dz, Athron:2007ry}.

The naturalness principle is not without its detractors. It has been pointed out that the anthropic principle could result in apparent violation of the naturalness principle in our observable universe \cite{Nelson}. This idea seems to have gained traction recently as the LHC has not revealed a solution to the hierarchy problem. It has also been argued that the naturalness principle is more an interesting historical and sociological factor in physics than a useful aid in objectively determining the truth value of theories \cite{Grinbaum:2009sk}.

At the same time, however, there is growing recognition that the naturalness principle may be rooted in Bayesian model comparison. The connection between the naturalness principle and Bayesian model comparison was noticed in Ref.~\citen{Strumia:1999fr}. In Ref.~\citen{Allanach:2007qk}, it was shown that Bayesian model comparison could be used to reach some of the same conclusions that previously had been reached by the naturalness principle. Then Refs.~\citen{Cabrera:2008tj,Cabrera:2009dm,Ghilencea:2012qk,Fichet:2012sn} used Bayesian model comparison to derive the BG sensitivity. Not only did this show that physicists' most popular quantitative measure of fine-tuning can be derived from Bayesian model comparison, but also Bayesian model comparison provides an objective interpretation of that fine-tuning measure. In Ref.~\citen{Fichet:2012sn} it is furthermore shown that the derived fine-tuning measure encompasses not only the BG sensitivity but also some of the refinements proposed in Refs.~\citen{Anderson:1994dz,Athron:2007ry}, and that it addresses some of the ambiguities in the BG sensitivity.

The Bayesian approach to naturalness has caught on and been put to use in recent years. It was applied to a number of MSSM-related model comparisons in Refs.~\citen{Fowlie:2014xha,Fowlie:2014faa,Fowlie:2015uga,Athron:2017fxj}. It has also been applied to more general extensions of the SM in Ref.~\citen{Clarke:2016jzm} and to the relaxion mechanism in Ref.~\citen{Fowlie:2016jlx}. In Ref.~\citen{AbdusSalam:2015uba}, Bayesian naturalness is used to argue that natural supersymmetry is still viable, and to identify the more natural allowed regions of parameter space. In Ref.~\citen{Kim:2013uxa} it is used to compare the CMSSM to the CNMSSM. In Ref.~\citen{Dumont:2013wma} it is used to study higher-dimensional operators in the Higgs sector. In Ref.~\citen{Ghilencea:2012gz} the Bayesian roots of the BG sensitivity are used to provide context and interpretation of an analysis based on BG sensitivity. Some of these developments have been discussed in Ref.~\citen{Ghilencea:2013fka}. Useful review can be found in Ref.~\citen{Kvellestad:2015cpa}.

It has also been noticed, though less talked about in the physics literature, that Bayesian model comparison accounts for the simplicity and testability of theories. See Ref.~\citen{Mackay} for an earlier paper on this and see Ref.~\citen{FowlieTalk} for a more recent discussion. See Refs.~\citen{Nesseris:2012cq,March,Kunz:2006mc} for some discussion of this in the context of evaluating cosmological models. See Refs.~\citen{Earman,Forster,Oppy} for discussion in the philosophy of science literature. In Ref.~\citen{Nesseris:2012cq} the connection to testability (predictiveness) is made explicit.

Bayesian naturalness has been developed in the context of discussing supersymmetry and the little hierarchy problem, and it has seen the most use within this context (see for example the references in the previous paragraph. See Refs.~\citen{Clarke:2016jzm,Fowlie:2016jlx} for two examples outside this context). In this paper, we apply Bayesian model comparison to gauge unification, an area to which it has not be applied before. Some of the Grand Unified Theories (GUTs) that we consider have additional parameters related to threshold corrections, making them more complicated than a simple GUT. Our analysis shows that despite this added complexity, these GUTs are preferable to the puzzle model because the observables are rather insensitive to the additional parameters.

The proton mass is important because applying the BG sensitivity leads to a paradox. The BG sensitivity seems to suggest that the proton mass is fine-tuned even though most physicists would agree that it is not actually fine-tuned\cite{Anderson:1994dz,Athron:2007ry}. We show that with Bayesian naturalness this paradox is resolved, thus justifying the claim that Bayesian naturalness is a more correct way to quantitatively understand naturalness. We will also apply the language of Bayesian naturalness to a simple GUT. Simplicity sits alongside naturalness as another intuitive, or aesthetic, criterion that physicists have used to guide research throughout history. The fact that Bayesian model comparison can quantitatively weigh both naturalness and simplicity has uses that we demonstrate by applying it to a number of more complicated GUTs, a task that requires original development of a Monte Carlo integration program utilizing the computational power of a Graphics Processing Unit (GPU).

The paper is organized as follows. In section \ref{sec:2} the Bayesian naturalness formalism is discussed, reviewing some existing literature on the topic. In section \ref{sec:3} we review the derivation of the BG sensitivity from Bayesian model comparison in a simple case of a model with one observable and one parameter. Then we apply Bayesian model comparison to the proton mass, and a number of GUTs. Section \ref{sec:4} presents our conclusions and discussion. \ref{sec:appendix} contains some brief details about how we implement Monte Carlo integration on a GPU. We have released our code to the public\cite{git}.

\section{Bayesian Model Comparison}
\label{sec:2}
\subsection{Bayes factors}

Bayes' theorem is a fundamental theorem in probability that follows deductively from the definition of conditional probability. In the context of making scientific inferences it can be written as
\begin{eqnarray}
p(\mathcal M|d)=p(d|\mathcal M)\frac{p(\mathcal M)}{p(d)}
\end{eqnarray}
where $\mathcal M$ is some model that is being considered, and $d$ is some experimental data. The quantity $p(\mathcal M|d)$ is called the posterior probability of model $\mathcal M$, and $p(\mathcal M)$ is called the prior probability.

There are two major problems with using Bayes' theorem to calculate posterior probabilities directly. First, such statements depend on the prior probabilities of all other hypothetical models that could predict data $d$. This enters Bayes' theorem through the quantity $p(d)$. Second, such statements will depend on the prior probability of the model being considered, $p(\mathcal M)$, which is subjective. The first of these two problems is avoided by considering the ratio of posterior probabilities of two models so that the prior probability of the data, $p(d)$ cancels out. Let us refer to the two models as $\mathcal P$ and $\mathcal M$ (the reason for this choice will become clear in subsection \ref{sec:puzzle}).
\begin{eqnarray}
\frac{p(\mathcal P|d)}{p(\mathcal M|d)}=\frac{p(d|\mathcal P)}{p(d|\mathcal M)}\frac{p(\mathcal P)}{p(\mathcal M)}\ .
\label{eq:bayesFactors}
\end{eqnarray}
The second problem is avoided by focusing on the quantity $p(d|\mathcal P)/p(d|\mathcal M)$. This quantity is referred to as the Bayes factor, $B_{\mathcal P\mathcal M}$.
\begin{eqnarray}
\frac{p(\mathcal P|d)}{p(\mathcal M|d)}&=&B_{\mathcal P\mathcal M}\frac{p(\mathcal P)}{p(\mathcal M)}\ ,\\
B_{\mathcal P\mathcal M}&=&\frac{p(d|\mathcal P)}{p(d|\mathcal M)}\ .
\end{eqnarray}
The Bayes factor quantifies whether the data $d$ favor model $\mathcal P$ over model $\mathcal M$ ($B_{\mathcal P\mathcal M} > 1$) and how strongly. The farther the Bayes factor is from unity, the more strongly the data $d$ favor model $\mathcal P$ over $\mathcal M$. In practice one is usually considering not the sum total of all relevant experimental data, but a specific experimental measurement or a set of experimental measurements. Referring to that data of immediate consideration as $d_2$, and all other baseline data as $d_1$, using Bayes' theorem, the ratio of posterior probabilities can then be written
\begin{eqnarray}
\frac{p(\mathcal P|d_2,d_1)}{p(\mathcal M|d_2,d_1)}&=&\frac{p(d_2|\mathcal P,d_1)}{p(d_2|\mathcal M,d_1)}\frac{p(\mathcal P|d_1)}{p(\mathcal M|d_1)}\nonumber\\
&=&\frac{p(d_2|\mathcal P,d_1)}{p(d_2|\mathcal M,d_1)}\frac{p(d_1|\mathcal P)}{p(d_1|\mathcal M)}\frac{p(\mathcal P)}{p(\mathcal M)}\nonumber\\ .
&=&B_{\mathcal P\mathcal M}^{(2)}B_{\mathcal P\mathcal M}^{(1)}\frac{p(\mathcal P)}{p(\mathcal M)}\ .
\end{eqnarray}
The $B_{\mathcal P\mathcal M}^{(2)}$ and $B_{\mathcal P\mathcal M}^{(1)}$ are called partial Bayes factors\cite{Fowlie:2014xha,Balazs:2013qva}. For brevity and compact notation, from here on in this paper, when we say ``Bayes factor'' we are referring to a partial Bayes factor. And we do not explicitly write the $^{(1)}$ or $^{(2)}$ superscripts or subscripts, nor do we explicitly include baseline data in our notation. It should be clear from context what data is being considered throughout this paper.

The numerator and denominator of the Bayes factor are referred to as the Bayesian evidence for $\mathcal P$ and $\mathcal M$, and are denoted $\mathcal Z_{\mathcal P}$ and $\mathcal Z_{\mathcal M}$ respectively so that the Bayes factor can be written
\begin{eqnarray}
B_{\mathcal P\mathcal M}=\frac{\mathcal Z_{\mathcal P}}{\mathcal Z_{\mathcal M}}
\end{eqnarray}
The Bayesian evidence for a model is simply the probability of the data $d$ assuming the model $\mathcal M$ is true.

The Bayes factor for some data $d$ is the crucial quantity for comparing two models with respect to how much the data favor one model over the other. Note that $B_{\mathcal P\mathcal M}=1/B_{\mathcal M\mathcal P}$. Interpretation of Bayes factors is sometimes aided by the Jeffreys' scale \cite{Fichet:2012sn,Fowlie:2014xha,Jeffreys}.

\subsection{Bayesian evidences}

Bayes factors are computed by computing the Bayesian evidences in the numerator and the denominator. Computing the Bayesian evidence for a model with $n$ unknown parameters, referred to as $\theta_i$ where $i = 1\cdots n$, requires assigning a prior probability distribution, $p(\theta_i)$, to the parameters. Suppressing the indices from here forward, the Bayesian evidence can then be written as
\begin{eqnarray}
\mathcal Z_{\mathcal M} = \int p(d|\theta)p(\theta|\mathcal M)d\theta\ ,
\end{eqnarray}
where the integral is over the entire $n$ dimensional parameter space and $d\theta$ denotes the volume element in parameter space. The probability $p(d|\theta)$ is referred to as a likelihood function. Many experimental results are Gaussian likelihood functions, due to the central limit theorem\cite{Fowlie:2014xha}. For example the experimental result that the $Z$-boson mass is $M_Z = 91.1876\pm0.0021\mbox{ GeV}$ \cite{Olive:2016xmw} means the likelihood $p(d|M_Z)$ is a Gaussian function of $M_Z$ with central value 91.1876 GeV and standard deviation 0.0021 GeV. The $m$ observables are referred to as $\mathcal O_i$ where $i=1\cdots m$. The Bayesian evidence can then be written as
\begin{eqnarray}
\mathcal Z_{\mathcal M} = \int p(d|\mathcal O)p(\theta|\mathcal M)d\theta\ ,
\end{eqnarray}
where the relationship between the observables and the parameters, $\mathcal O(\theta)$, in the model $\mathcal M$ is used to compute the integral. As long as there is no covariance between observables (or it can be neglected) the likelihood function $p(d|\mathcal O)$ can be written as a product of separate likelihood functions for each observable. If the experimental uncertainty is sufficiently small, these likelihood functions can be approximated by Dirac $\delta$-functions inside an integral. That is, using the form of a normalized Gaussian probability distribution with mean $\mu$ and variance $\sigma^2$,
\begin{eqnarray}
\int \frac{1}{\sigma\sqrt{2\pi}}e^{-\frac12\frac{(x-\mu)^2}{\sigma^2}}f(x)dx\approx\int \delta(x-\mu) f(x)dx\ ,
\end{eqnarray}
as long as $f(x)$ doesn't vary too much close to $x=\mu$. Both of these simplifications were used explicitly in Ref.~\citen{Fowlie:2014xha}, for example.

\subsection{Puzzle models}
\label{sec:puzzle}

Comparisons between two specific models, for example the CMSSM and the MSSM, are instructive. However, they do not capture the whole of the naturalness principle. It was shown in Ref.~\citen{Fichet:2012sn} that a puzzle model, referred to as $\mathcal P$ can aid in this. 

A puzzle model, $\mathcal P$, defined as a model in which the observables are simply considered to be fundamental parameters, is particularly useful for demonstrating the existence of the (big and little) hierarchy problem. Consider arguments that the CMSSM has a little hierarchy problem. To what model should the CMSSM be compared? Comparison to the SM is problematic because, depending on how one handles regularization of quadratic divergences, the SM may have a big hierarchy problem that dwarfs the CMSSM's little hierarchy problem \cite{Fowlie:2014xha}. Instead, the CMSSM should be compared to a puzzle model, defined such that the electroweak scale is a fundamental parameter of the model. Then the Bayes factor favors the puzzle model\cite{Fowlie:2014xha,Balazs:2013qva}.\footnote{Such a comparison is carried out in Refs.~\citen{Fowlie:2014xha,Balazs:2013qva} by comparing the CMSSM to the SM less quadratic divergences. It is pointed out in Ref.~\citen{Balazs:2013qva} that the latter is essentially a puzzle model.} The fact that Bayesian model comparison favors the puzzle model over the CMSSM is a manifestation of the little hierarchy problem. The Bayes factor can then serve as a quantitative measure of fine-tuning, like the BG sensitivity. The Bayes factor in Bayesian comparison with a puzzle model can be used to reveal a number of fine-tuning problems, including the big and little hierarchy problems, and the cosmological constant problem.

\subsection{Prior distributions}
\label{sec:priors}
We use the log-uniform prior for parameters throughout this paper:
\begin{eqnarray}
p(\theta)d\theta=\frac{1}{\log\frac{\theta_{\max}}{\theta_{\min}}}\frac1\theta d\theta\ .
\end{eqnarray}
Note that the log-uniform prior follows from requiring that the log of $\theta$ have a uniform distribution. The quantity $\log(\theta_{\max}/\theta_{\min})$ is called the prior volume. The log-uniform prior is not uncommon in the literature on Bayesian naturalness. See, for example, Refs.~\citen{Fichet:2012sn,Fowlie:2016jlx,Athron:2017fxj,Fowlie:2015uga,Fowlie:2014faa,Fowlie:2014xha}.

The justification is that it is invariant under power and scale transformations of the parameters. That is to say, if the parameter $\theta$ has a log-uniform prior, then $\theta^\prime=b\theta^a$ also has a log-uniform prior. Note that in the region between $\theta_{\min}$ and $\theta_{\max}$, the log-uniform prior is proportional to $1/\theta$. This is similar to certain cases of the noninformative Jeffrey's prior and reference priors that take the form $p(\theta)\propto 1/\theta$. See Ref.~\citen{Kass} for review.

This invariance is important for application to physics models. For example, the relevant observable in the hierarchy problem is the electroweak scale. The puzzle model will have the electroweak scale as a fundamental parameter with a log-uniform prior. But to which exact quantity should the log-uniform prior be assigned? Should it be $m_Z$, $m_Z^2$, or perhaps even $m_W$ or the Higgs VEV, $v$? With the log-uniform prior, all of these choices are equivalent because assigning the log-uniform prior to one means they all have a log-uniform prior. A similar question arises in gauge unification. Should the fundamental parameter in gauge unification models be taken to be the gauge coupling, $g$, or $\alpha=g^2/(4\pi)$? With the log-uniform prior, both choices are equivalent.

\section{Implications}
\label{sec:3}

\subsection{BG-sensitivity and the hierarchy problems}
\label{sec:bgsens}

The BG sensitivity arises in a straightforward way from the Bayesian formulation of naturalness. Derivation of the BG sensitivity from Bayesian naturalness has been done in Refs.~\citen{Cabrera:2008tj,Cabrera:2009dm,Ghilencea:2012qk,Fichet:2012sn}. Here, we review this derivation in a simple case and we use the result in the next subsection.

Consider a Bayesian comparison between a puzzle model, $\mathcal P$ and a candidate model, $\mathcal M$, with a parameter, $\theta$. The comparison will use a single observable $\mathcal O$ that has been measured in experiments and found to have value $\mathcal O_{ex}$ with very small experimental uncertainty. The likelihood function can then be approximated as a $\delta$ function:\footnote{It is not required that a likelihood function $p(d|\mathcal O)$ be normalized so that its integral with respect to $\mathcal O$ is unity. However, any overall coefficient will just cancel out in $B_{\mathcal P \mathcal M}$ so it is not necessary to account for the normalization here.} $p(d|\mathcal O)=\delta(\mathcal O-\mathcal O_{ex})$. As discussed in subsection \ref{sec:priors}, log-uniform priors should be used for both the observable and the parameter. The Bayesian evidence for the puzzle model then becomes
\begin{eqnarray}
\mathcal Z_{\mathcal P} = \frac{1}{\log\frac{\mathcal O_{\max}}{\mathcal O_{\min}}}\frac{1}{\mathcal O_{ex}}\ .
\end{eqnarray}
Assuming there is a unique point in the one-dimensional parameter space for which the observable takes its experimental value, the Bayesian evidence for the candidate model becomes
\begin{eqnarray}
\mathcal Z_{\mathcal M} = \frac{1}{\log\frac{\theta_{\max}}{\theta_{\min}}}\frac{1}{\theta_{ex}}\frac{1}{\left.\frac{\partial \mathcal O}{\partial \theta}\right|_{\theta_{ex}}}\ ,
\end{eqnarray}
where $\theta_{ex}$ denotes the value of $\theta$ for which $\mathcal O$ takes its experimental value. That is, $\mathcal O(\theta_{ex})=\mathcal O_{ex}$. The Bayes factor, $B_{\mathcal P \mathcal M}=\mathcal Z_{\mathcal P}/\mathcal Z_{\mathcal M}$, then becomes
\begin{eqnarray}
B_{\mathcal P \mathcal M}&=&\frac{\log\frac{\theta_{\max}}{\theta_{\min}}}{\log\frac{\mathcal O_{\max}}{\mathcal O_{\min}}} \left.\frac{\theta}{\mathcal O} \frac{\partial \mathcal O}{\partial \theta}\right|_{\theta_{ex}} \nonumber\\
&=&\frac{\log\frac{\theta_{\max}}{\theta_{\min}}}{\log\frac{\mathcal O_{\max}}{\mathcal O_{\min}}} \left.\frac{\partial \log \mathcal O}{\partial \log \theta}\right|_{\theta_{ex}}\ .
\label{eq:bgsens}
\end{eqnarray}
The partial derivative to the right, ${\partial \log \mathcal O}/{\partial \log \theta}|_{\theta_{ex}}$, is exactly the BG sensitivity. The ratio multiplying it is the ratio of prior volumes. A large BG sensitivity would imply that the puzzle model is favored and the candidate model disfavored, as long as the ratio of the prior volumes is not too different from unity.

Note the few assumptions necessary to demonstrate the connection between the Bayes factor and the BG sensitivity: similar prior volumes, log-uniform priors for both the observable and the parameter, and a sufficiently precise measurement of the observable so that the likelihood can be approximated by a $\delta$ function.

The observation that prior volumes affect the Bayes factor is not new. For example, Bartlett's paradox\cite{Bartlett} points out that when an alternative hypothesis uses an improper prior for a parameter to be estimated (for example a uniform prior over the interval $(-\infty,\infty)$ is considered), the posterior probability of the null hypothesis can become unity regardless of the data.

Some comments are in order regarding the ratio of prior volumes, which multiplies the BG sensitivity in equation \eqref{eq:bgsens}. Consider the case of the little hierarchy problem, which was the original motivation for the invention of the BG sensitivity. Here the observable, $\mathcal O$, is the electroweak scale. In the standard model, the electroweak scale is proportional to the mass term of the Higgs doublet. In the standard model without quadratic divergences, the mass of the Higgs doublet is a fundamental parameter. Therefore the standard model is equivalent to the puzzle model for the purposes of computing Bayes factors. The candidate model under consideration is the MSSM, where the fundamental parameter of interest to the little hierarchy problem is usually taken to be the $\mu$-term. The SM Higgs doublet mass and the MSSM $\mu$-term are of the same basic type. That is, they are both dimensionful mass terms not protected by any symmetry. Therefore it is reasonable to suppose them to have similar prior volumes. So the ratio of prior volumes is close to unity and the Bayes factor is approximately just the BG sensitivity. Not only does this explain why the BG sensitivity is appropriate for quantifying fine-tuning in the MSSM, but it also gives the BG sensitivity a concrete interpretation in terms of probabilities. Perhaps most importantly it suggests how we might identify situations in which the BG sensitivity is not appropriate for quantifying fine-tuning.

\subsection{Proton mass}

The mass of the proton can be estimated as the energy scale at which the strong coupling constant becomes non-perturbative. As noted in Refs.~\citen{Anderson:1994dz,Athron:2007ry}, this energy scale is very sensitive to the high-scale boundary value of the strong coupling.\footnote{``High scale'' is usually taken to be the Planck mass.} As a result, the proton mass has a very high BG sensitivity. This is a paradox if the BG sensitivity is taken as a measure of fine-tuning, since most physicists would agree that the proton mass is not fine-tuned.

In search for a resolution of this paradox, it is pointed out in Ref.~\citen{Anderson:1994dz} that one difference between the sensitivity in the proton mass and the sensitivity in some fine-tuning problems, such as the (big and little) hierarchy problem, is that the sensitivity in the proton mass is a global sensitivity. That is, the proton mass is highly sensitive over the entire parameter space. In the hierarchy problems, however, the electroweak scale is only highly sensitive in a small part of the parameter space and it is that specific part of the parameter space that nature has chosen. Motivated by this observation, the authors of Ref.~\citen{Anderson:1994dz} propose a new measure of fine tuning that is normalized by a kind of average sensitivity so that the proton mass, according to this new measure, is not fine-tuned.

In fact, according to Bayesian naturalness, the proton is not fine-tuned. This means it addresses the paradox noted in Ref.~\citen{Anderson:1994dz} automatically. The results of the previous section will help illuminate this. Following the analysis in Ref.~\citen{Anderson:1994dz}, using the one-loop renormalization group equation (RGE), the low-energy value of the strong coupling can be written as
\begin{eqnarray}
\alpha^{-1}_3(\mu)=\alpha^{-1}_3(M_{P})-\frac{b_3}{2\pi}\log\frac{\mu}{M_{P}}\ ,
\end{eqnarray}
where $M_P$ is the Planck mass. For this discussion of the proton mass we neglect any threshold corrections, as their impact on the numerical results would not be large enough to change any of our conclusions. Using the scale at which the strong coupling becomes unity as the proton mass, that is $\alpha^{-1}_3(m_p)=1$ where $m_p$ is the proton mass, yields
\begin{eqnarray}
1=\alpha^{-1}_3(M_{P})-\frac{b_3}{2\pi}\log\frac{m_p}{M_{P}}\ .
\label{eq:proton mass}
\end{eqnarray}
Treating $m_p$ as the observable and $\alpha_3(M_P)$ as a fundamental parameter. We can write an expression for the observable in terms of the parameter. Abbreviating $\alpha_3(M_P)$ as $\alpha_3$, 
\begin{eqnarray}
m_p(\alpha_3)=M_Pe^{\frac{2\pi}{b_3}(\alpha_3^{-1}-1)}\ .
\label{eq:func}
\end{eqnarray}
Then we can calculate the BG sensitivity.
\begin{eqnarray}
\frac{\partial \log m_p}{\partial \log \alpha_3}=-\frac{2\pi}{b_3}\alpha_3^{-1}\approx 45\ ,
\end{eqnarray}
where the numerical result is obtained by substituting $\alpha_3^{-1}\approx 50$ and $b_3=-7$\cite{Martin:1997ns}. If the BG sensitivity is taken as an indicator of fine tuning this would suggest that the proton mass is fine-tuned, a conclusion most physicists would intuitively disagree with. But when $B_{\mathcal P\mathcal M}$ is taken as the indicator of fine-tuning we see that the ratio of the prior volumes enters the calculation. Equation \eqref{eq:bgsens} can be used to calculate $B_{\mathcal P\mathcal M}$ in this case. 
\begin{eqnarray}
B_{\mathcal P\mathcal M} = \frac{\log\frac{\alpha_{3_{\max}}}{\alpha_{3_{\min}}}}{\log\frac{m_{p_{\max}}}{m_{p_{\min}}}}\left(-\frac{2\pi}{b_3}\right)\alpha_3^{-1}\ .
\end{eqnarray}
Unlike in the case of the little hierarchy problem, there is no reason to suppose that the ratio of prior volumes should be close to unity. The observable, $m_p$, is a dimensionful mass while $\alpha_3$ is a dimensionless coupling constant, so there is no reason they should have similar prior volumes. Therefore a large BG sensitivity should not be taken to imply that $B_{\mathcal P\mathcal M}$ is large and the proton mass fine-tuned. 

Not only is there no reason to suppose that the ratio of the prior volumes is close to unity, but such an supposition seems biased toward favoring the puzzle model and finding the proton mass to be fine-tuned. Suppose that the prior for $\alpha_3$ is chosen to cover the range from 0.001 to 1, or three orders of magnitude. We then find from equation \eqref{eq:func} that the range of values of $m_p$ that could be obtained from the model spans approximately 390 orders of magnitude. This is due to the large global sensitivity noted in Ref.~\citen{Anderson:1994dz}. Supposing that the prior volume of $m_p$ should be comparable to the prior volume of $\alpha_3$ (three orders of magnitude in this case) arbitrarily restricts the puzzle model to a relatively narrow interval around the observed value of $m_p$ while the candidate model is allowed to span a much larger interval. This, unsurprisingly, would make the Bayes factor appear to favor the puzzle model.

Perhaps it would be sufficient to leave the discussion of the proton mass here. The reader is hopefully convinced that there is no reason the ratio of prior volumes should be close to unity, and thus no reason to take the BG sensitivity at face value as an indicator of fine-tuning in the proton mass. Furthermore, assuming the ratio of prior volumes to be close to unity is biased toward finding the proton mass to be fine-tuned. Given that such an assumption is implicit in using the BG sensitivity as an indicator of fine-tuning (see subsection \ref{sec:bgsens}), it is no surprise that the BG sensitivity makes the proton mass appear fine-tuned. Thus the paradoxical result of applying the BG sensitivity to the proton mass is resolved. According to Bayesian naturalness, the proton mass is not fine-tuned and there is no reason to take the BG sensitivity as an indicator of fine-tuning in the proton mass. That said, the discussion up to this point suggests a possible heuristic for restricting the ratio of prior volumes, thereby enabling us to make some more concrete statements about $B_{\mathcal P\mathcal M}$. This heuristic, which we call the fairness heuristic, is not really necessary given the above discussion, but it is interesting to explore its implications.

The fairness heuristic is stated as follows: the prior volumes for the puzzle model and the candidate model should be chosen so that the two models are allowed to cover the same range of values for the observable. We refer to this heuristic simply as the fairness heuristic because it enforces a kind of fairness in choosing prior volumes for the two models. Applied to the case of the proton mass, the fairness heuristic says simply
\begin{eqnarray}
m_{p_{\max}}&=&m_p(\alpha_{3_{\max}})\nonumber\\
m_{p_{\min}}&=&m_p(\alpha_{3_{\min}})\ .
\label{eq:fairness}
\end{eqnarray}
An appropriate choice for $\alpha_{3_{\max}}$ is unity, because if it were any larger the theory would be non-perturbative. Applying the fairness heuristic using equation \eqref{eq:func} yields
\begin{eqnarray}
\log\frac{m_{p_{\max}}}{m_{p_{\min}}}=\frac{2\pi}{b_3}(\alpha_{3_{\max}}^{-1}-\alpha_{3_{\min}}^{-1})\approx-\frac{2\pi}{b_3}\alpha_{3_{\min}}^{-1}\ .
\end{eqnarray}
The $B_{\mathcal P\mathcal M}$ then becomes
\begin{eqnarray}
B_{\mathcal P\mathcal M} \approx -\ln(\alpha_{3_{\min}})\alpha_{3_{\min}}\alpha_3^{-1}\lesssim 4\ ,
\label{eq:protonBayesFactor}
\end{eqnarray}
where the numerical result is obtained using the fact that $\alpha_{3_{\min}}$ must be less than the measured value $\alpha_3\approx 1/50$.

A few comments are in order, as there are strong grounds for questioning the fairness heuristic. First, when comparing more than two models, as we do below, it is not clear which of multiple candidate models should be used to determine the prior volume of the puzzle model. Second, the fairness heuristic is only meaningful when the prior distributions have some maximal and minimal values for the parameters. Prior distributions which cover the interval $(-\infty,\infty)$ have no such maximal and minimal values and the fairness criterion is meaningless there. Third, even if the prior has maximal and minimal values, In some cases these may not correspond to any maximal and minimal values of the observables. In such cases, there is no way to apply the fairness heuristic. Finally, a model that predicts that an observable should lie in a narrow range ought to be strongly favored, but applying the fairness heuristic in such a case would result in a low prior volume for the puzzle model, favoring the puzzle model instead of the highly predictive candidate model. In light of these points, we do not claim that the fairness heuristic should be used for model comparisons beyond this particular example. We emphasize again that the fairness heuristic is not really necessary to reach our central conclusion in this subsection, which is that, according to Bayesian naturalness, the proton mass is not necessarily fine-tuned and the BG sensitivity should not be taken at face value as an indicator of fine-tuning in the proton mass.
\subsection{The weak-scale MSSM GUT}
\label{sec:simplegut}

In previous subsections we have discussed Bayesian naturalness as it relates to the (big and little) hierarchy problem, and to the proton mass. These problems are not unfamiliar to the literature on naturalness, and it is increasingly recognized that Bayesian naturalness reliably reproduces physicists' intuitive notion of naturalness. As we mentioned in the introduction, it has also been noticed that Bayesian model comparison reproduces physicists' notion of simplicity\cite{Mackay,FowlieTalk,Nesseris:2012cq,March,Kunz:2006mc,Earman,Forster,Oppy}. We demonstrate an example by applying Bayesian model comparison to the weak-scale MSSM GUT\cite{Martin:1997ns,Dimopoulos:1981zb,Langacker:1990jh,Ellis:1990wk,Amaldi:1991cn,Langacker:1991an,Giunti:1991ta} (see Ref.~\citen{Martin:1997ns} and references therein for review). This will also be a useful preliminary to applying Bayesian model comparison to the more complicated GUTs below.

\begin{figure}
\center
\includegraphics{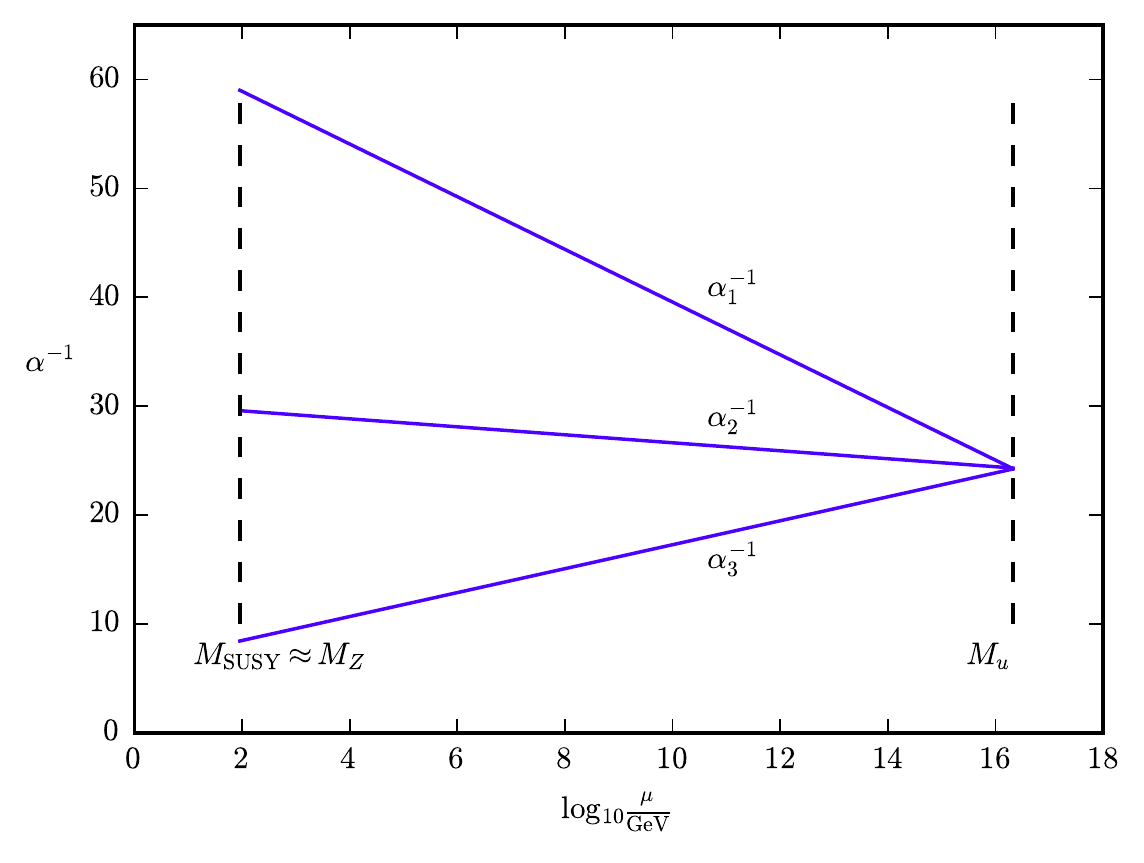}
\caption{\small Gauge coupling unification in the weak-scale MSSM GUT with $M_{\susy}\approx M_{Z}$. Renormalization scale is denoted $\mu$. This plot was created using the values $M_\susy=M_Z$ and $M_u=2.1\times 10^{16}\mbox{ GeV}$, along with the experimental values and slope factors given in the text.}
\label{fig:simpleGut}
\end{figure}

The MSSM contains the SM matter fields each with scalar superpartners, the SM gauge bosons each with fermionic superpartners (gauginos), and two Higgs supermultiplets to effect electroweak symmetry breaking. One important consequence of the additional particle content of the MSSM is that it changes the slope factors in the renormalization group equations of the gauge couplings. When the gauge couplings are evolved to high-scale under the MSSM slope factors, they unify almost exactly (i.e. within experimental uncertainty). This unification is shown in Fig. \ref{fig:simpleGut}.

Supersymmetry in the MSSM is broken via soft mass parameters, which give mass to the scalar superpartners and the gauginos above the masses of their SM counterparts. The soft mass parameters are typically assumed to be all around the same mass scale, called $M_\susy$. At mass scales much less than $M_\susy$, the scalar superpartners and gauginos are decoupled and the particle content and slope factors are effectively those of the SM. The parameter $M_\susy$ is then important to unification because it defines the scale at which the slope factors change from their SM to the MSSM values.

The weak-scale MSSM GUT assumes that the SUSY scale, $M_\susy$ is approximately equal to the electroweak scale, $M_Z$. LHC searches have placed a lower bound on sparticle masses that requires that the SUSY scale be substantially higher. Below we add threshold corrections to this model so that the SUSY scale can be higher. First we discuss this weak-scale MSSM as it serves as an instructive example of how Bayesian model comparison reproduces conclusions that otherwise would be rooted in an aesthetic principle that simple theories are to be favored.

The weak-scale MSSM GUT is favored on the basis of simplicity because it can explain the measured values of the three gauge couplings $\alpha_1$, $\alpha_2$, and $\alpha_3$, using fewer than three parameters. The intuitive conclusion is that the GUT is favorable because having fewer parameters is simpler. In this subsection we use Bayesian model comparison to arrive at the same conclusion.

We choose the two parameters of the GUT to be the unified gauge coupling, $\alpha_u$, and the scale of unification, $M_u$. The SUSY scale, $M_{\susy}$, representing the average mass of the sparticles, is assumed to be approximately the electroweak scale, that is, $M_{\susy}\approx M_Z$. We work with log-uniform priors, so our results are independent of scaling or power law redefinitions of the parameters, as shown in subsection \ref{sec:priors}.

The non-unifying model treats each of the gauge couplings each as separate fundamental parameters, so it is a puzzle model. In the GUT, one-loop RGEs relate the observables to the parameters.
\begin{eqnarray}
\alpha_a^{-1}=\alpha_u^{-1}+\frac{b_a}{2\pi}\ln\frac{M_u}{M_Z}
\label{eq:gaugeRunning}
\end{eqnarray}
where $a\in\{1,2,3\}$. The Bayes factor favoring the GUT, $\mathcal M$, over the non-unifying model, $\mathcal P$, is
\begin{eqnarray}
B_{\mathcal M \mathcal P}&=&\frac{\mathcal Z_{\mathcal M}}{\mathcal Z_{\mathcal P}}\ ,
\end{eqnarray}
where,
\begin{eqnarray}
\mathcal Z_{\mathcal M}&=&\int p(d|\mathcal O)p(\theta|\mathcal M)d\theta\nonumber\\
\mathcal Z_{\mathcal P}&=&\int p(d|\mathcal O)p(\mathcal O|\mathcal P)d\mathcal O\ .
\end{eqnarray}
The likelihood function is a product of three Gaussian likelihood functions, one for each of the three gauge couplings.

The Bayesian evidence for the puzzle model is an integral over a three-dimensional parameter space, and each of the three Gaussian factors in the likelihood function may be approximated as a $\delta$-function to evaluate the integral. Assuming the same prior for all three gauge couplings, that is,
\begin{eqnarray}
\alpha_{1_{\max}}=\alpha_{2_{\max}}=\alpha_{3_{\max}}\equiv\alpha_{\max}\nonumber\\
\alpha_{1_{\min}}=\alpha_{2_{\min}}=\alpha_{3_{\min}}\equiv\alpha_{\min}\ ,
\end{eqnarray}
and using the experimental values\cite{Olive:2016xmw},
\begin{eqnarray}
\alpha_1 = 0.016946 \quad \alpha_2 = 0.033793 \quad \alpha_3 = 0.1181
\end{eqnarray}
yields
\begin{eqnarray}
\mathcal Z_{\mathcal P} = \frac{1}{\ln^3\frac{\alpha_{\max}}{\alpha_{\min}}}\frac{1}{\alpha_1\alpha_2\alpha_3}\approx 124\ ,
\label{eq:zpuzzle}
\end{eqnarray}
where we have chosen $\alpha_{u_{\max}}$ to be unity, chosen $\alpha_{u_{\min}}$ to be the fine-structure constant ($\approx 1/137$).

The Bayesian evidence for the GUT is an integral over a two-dimensional parameter space. The Gaussian likelihood functions for $\alpha_1$ and $\alpha_2$ can be approximated as $\delta$-functions because their experimental uncertainties, $\sigma_1$ and $\sigma_2$, are much smaller than $\sigma_3$, the experimental uncertainty in $\alpha_3$. The experimental uncertainties are\cite{Olive:2016xmw}
\begin{eqnarray}
\sigma_1 = 3.5\times 10^{-6} \quad \sigma_2 = 1.9\times 10^{-5} \quad \sigma_3 = 0.0011
\end{eqnarray}
These $\delta$-functions can be used to evaluate the two-dimensional integral. Note that this necessitates a change of variables, yielding a Jacobian factor:
\begin{eqnarray}
\left|\det\left(\frac{\partial \mathcal O_i}{\partial \mathcal \theta_j}\right)\right|=\frac{1}{2\pi}\frac{\alpha_1^2\alpha_2^2}{\alpha_u^2M_u}\left|b_1-b_2\right|\ .
\end{eqnarray}
where $\mathcal O_i = (\alpha_1,\alpha_2)$ and $\theta_j = (\alpha_u,M_u)$. Using a normalized Gaussian likelihood for $\alpha_3$, the integral evaluates to
\begin{eqnarray}
\mathcal Z_{\mathcal M}&=&\frac1{\sigma_3\sqrt{2\pi}}e^{-\frac12\frac{(\alpha_3-\alpha_{3_{ex}})^2}{\sigma_3^2}}\frac{1}{\ln\frac{\alpha_{\max}}{\alpha_{\min}}}\frac1{\ln\frac{M_{u_{\max}}}{M_{u_{\min}}}}\frac{2\pi\alpha_u}{\alpha_1^2\alpha_2^2|b_1-b_2|}\nonumber\\
&\approx& 1.00\times 10^5\ ,
\label{eq:zgut}
\end{eqnarray}
where we have used $\sigma_3 = 0.0011$\cite{Olive:2016xmw}, chosen $M_{u_{\max}}$ to be the Planck scale, $M_P=1.22\times10^{19}$ GeV, chosen $M_{u_{\min}}$ to be the $Z$ mass, 91.1876 GeV,\cite{Olive:2016xmw} chosen $\alpha_{u_{\max}}$ to be unity, chosen $\alpha_{u_{\min}}$ to be the fine-structure constant ($\approx 1/137$), and used the well-known MSSM slope factors,
\begin{eqnarray}
b_3^\mssm = -3 \quad b_2^\mssm = 1 \quad b_1^\mssm = \frac{33}{5}\ .
\label{eq:mssmSlopeFactors}
\end{eqnarray}
The minimal and maximal values of the parameters are summarized in Table \ref{tab:simpleGutPriors}.

\begin{table}
\center
\begin{tabular}{|p{1.5in}|p{1.4in}|p{1.5in}|}
\hline
parameter & min & max\\
\hline
\hline
$M_u$ & $M_Z$ & $M_P$\\
\hline
$\alpha_u$ & 1/137 & 1\\
\hline
\end{tabular}
\caption{\small Minimal and maximal values of parameters used to compute the Bayesian evidence for the weak-scale MSSM GUT.}
\label{tab:simpleGutPriors}
\end{table}
Note that there is a unique value for $\alpha_u$ and $M_u$ that yields the experimentally observed values of $\alpha_1$ and $\alpha_2$ when substituted into equation \eqref{eq:gaugeRunning}. In equation \eqref{eq:zgut}, $\alpha_u$ refers to that unique value. This constraint was imposed by the two $\delta$-functions that were used to evaluate the integral. Furthermore, $\alpha_3$ in equation \eqref{eq:zgut} refers to the the unique value that is yielded by substituting those unique values of $\alpha_u$ and $M_u$ into equation \eqref{eq:gaugeRunning}. In general, this value need not match, or even be close to $\alpha_{3_{ex}}$. The fact that it does match (within experimental uncertainty) is what makes this unification model so appealing. It is also what makes the normalized Gaussian factor in the front of equation \eqref{eq:zgut} quite large, so that the conclusion from Bayesian model comparison echoes the physicist's intuitive conclusion based on simplicity.

The approximate analytical result in equation \eqref{eq:zgut} is attainable in the weak-scale MSSM GUT. But the GUTs discussed below are more complicated so all the Bayesian evidences computed below are computed using Monte Carlo integration run on a GPU. For a few brief details on the implementation, see \ref{sec:appendix}. We have released our code to the public\cite{git}. Monte Carlo integration was also used to verify the result in \eqref{eq:zgut}, yielding $\mathcal Z = (1.0569\pm.0014)\times 10^5$ where the uncertainty given here and in our other results throughout this paper is the $1\sigma$ statistical uncertainty arising from the Monte Carlo integration. The $\sim 6\%$ discrepancy is evidently not due to the Monte Carlo integration and is probably due to approximating the Gaussian likelihood functions as $\delta$-functions.

Taking the quotient of the Bayesian evidences to compute the Bayes factor yields
\begin{eqnarray}
B_{\mathcal M\mathcal P} \approx 806\ , 
\end{eqnarray}
so the Bayes factor strongly favors the weak-scale MSSM GUT over the puzzle model. This result is exactly in line with physicists' intuitive conclusion that the GUT is more favorable based on simplicity. So how has simplicity manifested itself in the Bayesian analysis? As mentioned earlier, due to the GUT having fewer free parameters than the puzzle model (a characteristic of simple theories), its prior parameter space is more restricted. Since this restricted parameter space is consistent with the observed data, it is much more probable that a random point selected from this restricted parameter space is consistent with the data than a random point selected from the much broader and less restricted parameter space of the puzzle model. Put another way, if the puzzle model were true, it would be quite surprising and a coincidence that the observed data happen to be consistent with unification, but if the GUT is true then it is not a coincidence but rather is to be expected. This probabilistic language is not unfamiliar to physicists discussing such matters as gauge unification, and it illuminates why the Bayesian analysis favors the simpler theory.

The more precise the measurement of the observables, the greater the coincidence and the more strongly Bayesian model comparison favors the simple model. This fact is reflected in the Bayesian analysis by the experimental uncertainty, $\sigma_3$, appearing in the denominator of the Gaussian factor in equation \eqref{eq:zgut}. Of course, if the observed data did not agree with the weak-scale MSSM GUT, that is, if $\alpha_3$ was different from $\alpha_{3_{ex}}$ by much more than the experimental uncertainty, the Gaussian factor in equation \eqref{eq:zgut} would rapidly go to zero, bringing the credibility of the GUT with it.

\subsection{The MSSM GUT}
The weak-scale MSSM GUT serves as a concrete example of Bayesian model comparison favoring a simple theory. However, the non-observation of sparticles at the LHC implies that the SUSY scale, if it exists, must be well above the electroweak scale. Accommodating this within an MSSM GUT requires modeling threshold corrections, going beyond the weak-scale MSSM GUT. One way to model the SUSY threshold corrections is to consider the effect of the colored superpartners decoupling at a higher mass scale than the non-colored superpartners.\footnote{This model of SUSY threshold corrections was studied in Ref.~\citen{Ovrut:2012wg} in the context of the $B-L$ MSSM.} This splits the mass scale $M_\susy$ into two mass scales: $M_{\susy_c}$ where the colored superpartners decouple, and $M_{\susy_n}$ where the non-colored superpartners decouple. Such a model is theoretically well motivated, because the gluino is pushed to a higher mass than the other gauginos by one-loop corrections involving the strong coupling constant. The squarks are in turn pushed to a higher mass than the other scalar superpartners by one-loop corrections involving the gluino. See Refs.~\citen{Martin:1997ns,Ovrut:2012wg,Ovrut:2015uea,Deen:2016vyh} for some discussion and relevant results. The result is that the colored superpartners on average decouple at a significantly higher mass scale than the non-colored superpartners.

\begin{figure}
\center
\includegraphics{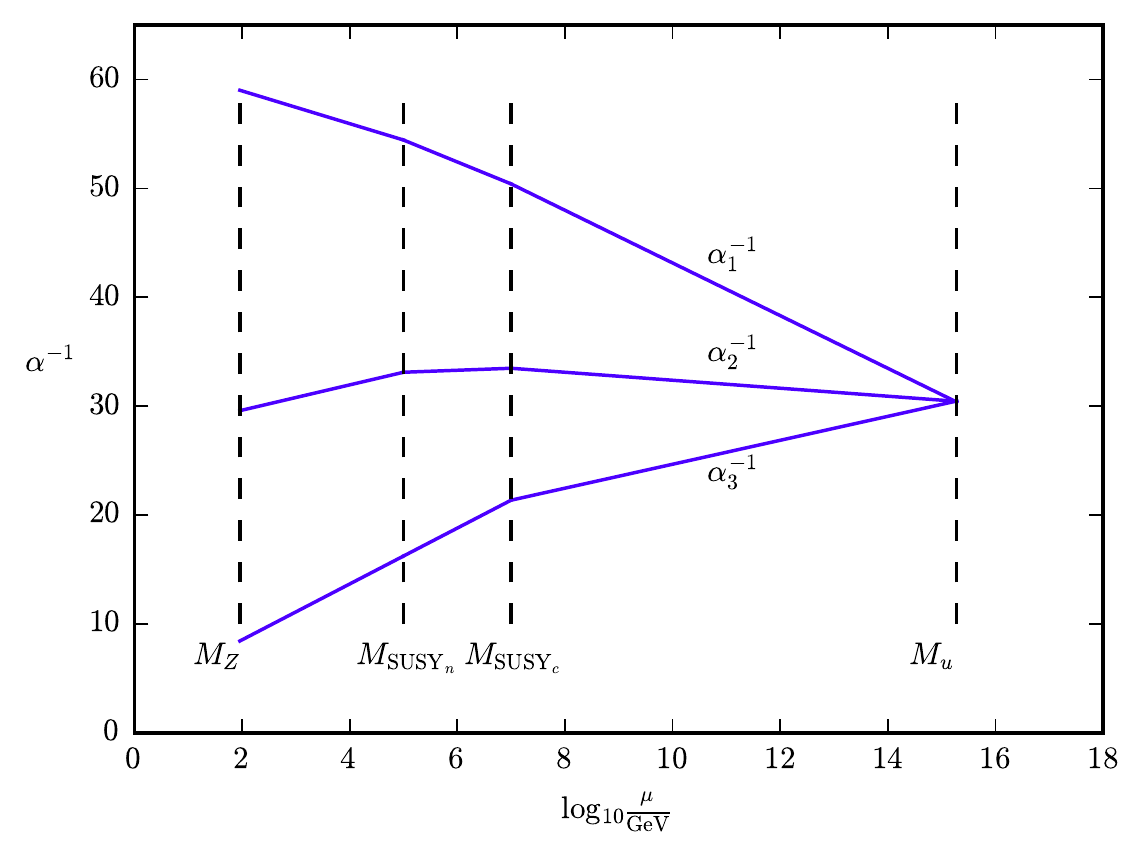}
\caption{\small Gauge coupling unification in the MSSM with colored superpartners decoupling at a significantly higher scale than the non-colored superpartners. Renormalization scale is denoted $\mu$. This plot was created using the values $M_{\susy_n}=10^5\mbox{ GeV}$, $M_{\susy_c}=10^7\mbox{ GeV}$, and $M_u=1.9\times 10^{15}\mbox{ GeV}$, along with the experimental values and slope factors given in the text. The scales $M_{\susy_n}$ and $M_{\susy_c}$ being separated by a factor of 100 may be unlikely, but these values are chosen for this plot to make their separation and the kinks in the plot more visually apparent.}
\label{fig:susyThresholdGut}
\end{figure}

The slope factors above the mass scales $M_{\susy_c}$ and $M_{\susy_n}$ are the MSSM slope factors given in equation \eqref{eq:mssmSlopeFactors}. Between the mass scales $M_{\susy_c}$ and $M_{\susy_n}$ the slope factors are called $b^\ncmssm_a$. They are \cite{Ovrut:2012wg}
\begin{eqnarray}
b^\ncmssm_3 = -7 \quad b_2^\ncmssm = -\frac12 \quad b_1^\ncmssm = \frac{11}{2}\ .
\label{eq:ncMssmSlopeFactors}
\end{eqnarray}
Below those scales the slope factors are the well-known SM slope factors:
\begin{eqnarray}
b_3^\sm = -7 \quad b_2^\sm = -\frac{19}{16} \quad b_1^\sm = \frac{41}{10}\ .
\label{eq:smSlopeFactors}
\end{eqnarray}
The relationship between the observables and the parameters of this model is then
\begin{eqnarray}
\alpha_{a}^{-1}&=&\alpha_u^{-1}+\frac{b_{a}^\mssm}{2\pi}\ln\frac{M_u}{M_{\susy_c}}+\frac{b_{a}^\ncmssm}{2\pi}\ln\frac{M_{\susy_c}}{M_{\susy_n}}+\frac{b_{a}^\sm}{2\pi}\ln\frac{M_{\susy_n}}{M_Z}\ ,
\label{eq:mssmSusyThresholdRelationship}
\end{eqnarray}
where $a\in\{1,2,3\}$. Splitting the SUSY scale in this theoretically motivated way turns out to provide just the right correction to allow the gauge couplings to unify, even with the SUSY scales well above the electroweak scale. This unification is shown in Fig. \ref{fig:susyThresholdGut}.

Viewing all of these scales as free parameters, one could argue that this GUT lacks simplicity because of the additional free parameters it introduces. It actually has more parameters than the puzzle model (four parameters, $\alpha_u$, $M_{\susy_n}$, $M_{\susy_C}$, and $M_u$, as opposed to three parameters $\alpha_3$, $\alpha_2$, and $\alpha_1$). On the other hand, the observables depend only logarithmically on these parameters, so the observables are not very sensitive to changes in these parameters. This means that the model lacks fine-tuning or, equivalently, is natural. When a model is natural, but not simple, how should these conflicting judgments be weighed? Bayesian model comparison gives a quantitative means to weigh the conflicting intuitive judgements of naturalness and lack of simplicity. 

In order to compute the Bayesian evidence for this unification model, we need priors for the four parameters $M_{\susy_c}$, $M_{\susy_n}$, $M_u$, and $\alpha_u$. There are a few reasonable choices. Perhaps the most obvious choice is to use the same prior for all of the mass scales. This prior, which we will refer to as the unconstrained SUSY threshold prior, is summarized in Table \ref{tab:susyThresholdGut1}.
\begin{table}
\center
\begin{tabular}{|p{1.5in}|p{1.4in}|p{1.5in}|}
\hline
parameter & min & max\\
\hline
\hline
$M_u$ & $M_Z$ & $M_P$\\
\hline
$M_{\susy_c}$ & $M_Z$ & $M_P$\\
\hline
$M_{\susy_n}$ & $M_Z$ & $M_P$\\
\hline
$\alpha_u$ & 1/137 & 1\\
\hline
\end{tabular}
\caption{\small Minimal and maximal values of parameters used to compute the Bayesian evidence for the weak-scale MSSM GUT.}
\label{tab:susyThresholdGut1}
\end{table}
This choice is unrealistic, however, because it allows the two SUSY scales to be drastically separated, while typically one would expect all sparticles to have masses mildly scattered around a single mass scale. The results in Refs.~\citen{Martin:1997ns,Ovrut:2012wg,Ovrut:2015uea,Deen:2016vyh} suggest a mild separation less than a factor of ten. We therefore introduce a parameter $h$ defined by
\begin{eqnarray}
M_{\susy_c}=hM_{\susy_n}\ ,
\end{eqnarray}
and assign to it a log-uniform prior with maximal and minimal values. Then $M_{\susy_n}$ is no longer treated as a fundamental parameter, it is calculated from $M_{\susy_c}$ and $h$, which are treated as fundamental parameters. We refer to this prior as the constrained SUSY threshold prior. The minimal value, $h_{\min}$, is always set to unity, corresponding to no SUSY threshold correction. We consider $h_{\max}$ values of 10, 100, and 1000. This is summarized in Table \ref{tab:constrainedSusyThresholdPriors}.
\begin{table}
\center
\begin{tabular}{|p{1.5in}|p{1.4in}|p{1.5in}|}
\hline
parameter & min & max\\
\hline
\hline
$M_u$ & $M_Z$ & $M_P$\\
\hline
$M_{\susy_c}$ & $M_Z$ & $M_P$\\
\hline
$h$ & 1 & 10, 100, or 1000\\
\hline
$\alpha_u$ & 1/137 & 1\\
\hline
\end{tabular}
\caption{\small Minimal and maximal values of parameters used to compute the Bayesian evidence for the MSSM GUT with the constrained SUSY threshold prior.}
\label{tab:constrainedSusyThresholdPriors}
\end{table}

We compute the Bayesian evidences for the MSSM GUT with both unconstrained and constrained SUSY threshold priors. Step functions are used in the likelihood to enforce $M_Z \leq M_{\susy_n} \leq M_{\susy_c} \leq M_u$. The results are given in Table \ref{tab:susyThresholdGutEvidences}.
\begin{table}
\center
\begin{tabular}{|p{3.2in}|p{1.3in}|}
\hline
model & Bayesian evidence, $\mathcal{Z}$\\
\hline
\hline
MSSM GUT with unconstrained threshold & $1204\pm19$\\
\hline
MSSM GUT with constrained threshold, $h_{\max}=10$ & $5316\pm40$\\
\hline
MSSM GUT with constrained threshold, $h_{\max}=100$ & $5189\pm40$\\
\hline
MSSM GUT with constrained threshold, $h_{\max}=1000$ & $5058\pm39$\\
\hline
\end{tabular}
\caption{\small Bayesian evidences for the MSSM GUT with unconstrained and constrained SUSY threshold priors.}
\label{tab:susyThresholdGutEvidences}
\end{table}
The MSSM GUT with SUSY thresholds is favored over the puzzle model ($\mathcal Z = 124$). This is true even though it is a more complicated GUT. Additional scales (parameters) are needed to make this GUT agree with experimental data making it more complicated than even the puzzle model. However, importantly, the observables are relatively insensitive to these new scales, meaning that the GUT is natural. Bayesian model comparison automatically accounts for this and weighs it against the lack of simplicity. This example demonstrates that Bayesian model comparison gives a quantitative way to weigh the intuitive notions of simplicity and naturalness, which is especially useful in cases like this where naturalness and simplicity seem to be in conflict and intuitive considerations alone cannot judge which of those criteria should be weighed more heavily.

Also of interest is the fact that the constrained prior, which is better motivated anyway, is more favored than the unconstrained prior. Even though the Bayesian evidence does not depend strongly on the value of $h_{\max}$. We regard $h_{\max}=10$ as the most well-justified value because it is suggested by the results in Refs.~\citen{Martin:1997ns,Ovrut:2012wg,Ovrut:2015uea,Deen:2016vyh}.

\begin{figure}
\begin{centering}
\center
\includegraphics{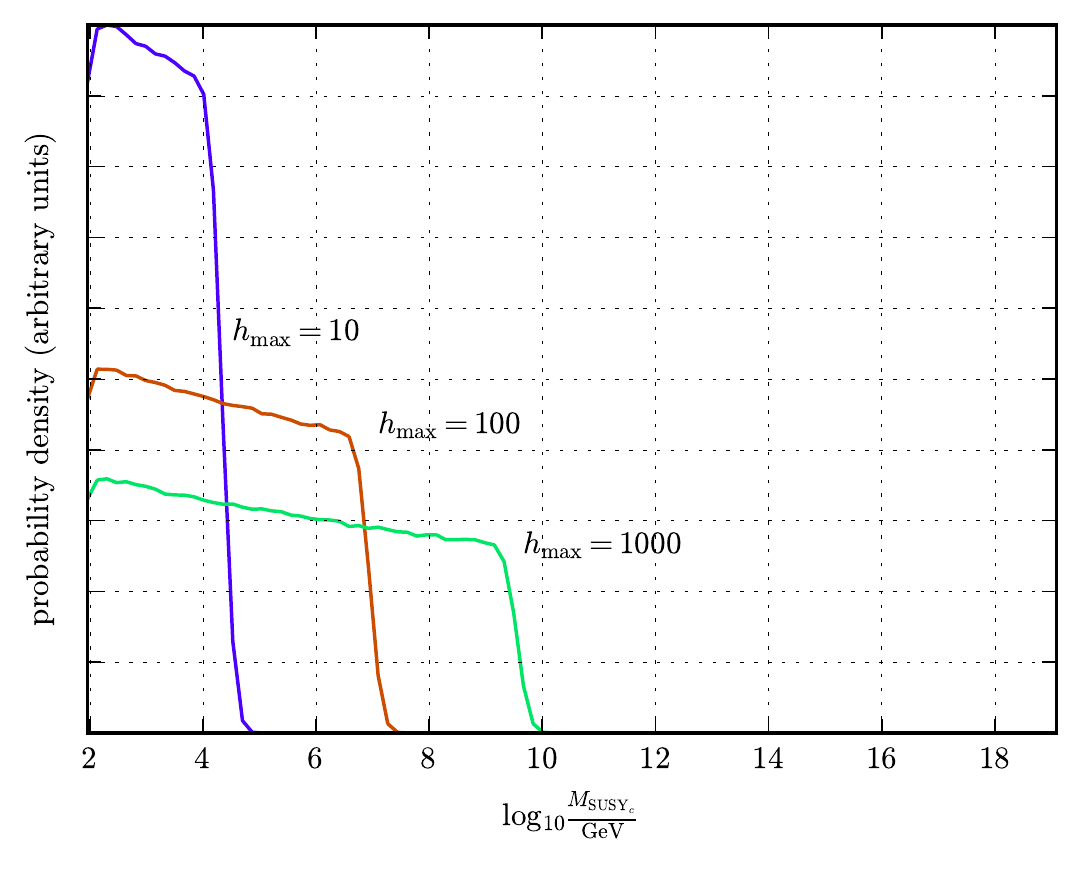}
\caption{\small Posterior probability distribution for $\log M_{\susy_c}$ in arbitrary units. The small fluctuations in the lines are due to the statistical uncertainty in the Monte Carlo integration. The three curves are normalized relative to each other so that they all have the same area underneath.}
\label{fig:posteriorSusyThreshold}
\end{centering}
\end{figure}

Fig. \ref{fig:posteriorSusyThreshold} shows the posterior probability distribution of $\log M_{\susy_c}$ in the case of the constrained SUSY threshold priors. The posterior probability distribution is given by Bayes theorem. That is,
\begin{eqnarray}
p(\log M_{\susy_c}|d) = p(d|\log M_{\susy_c})\frac{p(\log M_{\susy_c})}{p(d)}\ ,
\end{eqnarray}
where all probabilities are in the MSSM GUT with constrained SUSY threshold priors. Since $M_{\susy_c}$ is a fundamental parameter with a log-uniform prior, $p(\log M_{\susy_c})$ is constant, as is $p(d)$. We use Monte Carlo integration to compute $p(d|\log M_{\susy_c})$. The result is plotted in arbitrary units. This figure is limited to the constrained SUSY threshold priors because they are better motivated and favored by the Bayesian model comparison. The figure shows that the SUSY scale is not constrained to be weak-scale. It can be orders of magnitude higher and still be consistent with unification. However, exactly how much higher the SUSY scale can be depends on the prior chosen for the SUSY threshold. With $h_{\max} = 10$ it can be up to around 10 TeV, above the current LHC bounds. With $h_{\max} = 1000$, it can be as high as $10^9$ GeV.

In addition to threshold corrections at the SUSY scale, GUT models may also account for threshold corrections at the unification scale. See Refs.~\citen{Dienes:1996du,Deen:2016vyh,Deen:2016zfr,Kaplunovsky:1992vs,Mayr:1993kn,Dolan:1992nf,Kiritsis:1996dn,Klaput:2010dg,deAlwis:2012bm,Bailin:2014nna} for examples and discussion in the context of string theory. One way to do this is by introducing new parameters, $\Delta_1$, $\Delta_2$, and $\Delta_3$, which contain threshold corrections to the three gauge couplings. See, for example, Refs.~\citen{Langacker:1992rq,Deen:2016vyh}. The relationship between the parameters and observables is modified from equation \eqref{eq:mssmSusyThresholdRelationship} to 
\begin{eqnarray}
\alpha_{a}^{-1}&=&\alpha_u^{-1} + \frac{\Delta_a}{4\pi}\nonumber\\
&&+\frac{b_{a}^\mssm}{2\pi}\ln\frac{M_u}{M_{\susy_c}}+\frac{b_{a}^\ncmssm}{2\pi}\ln\frac{M_{\susy_c}}{M_{\susy_n}}+\frac{b_{a}^\sm}{2\pi}\ln\frac{M_{\susy_n}}{M_Z}\ ,
\label{eq:mssmSusyThresholdNuisanceRelationship}
\end{eqnarray}
where $a\in\{1,2,3\}$. We assign to $|\Delta_a|$ a log-uniform prior with minimal value of $\Delta_{\min}=1$ and consider maximal values of $\Delta_{\max} =10\mbox{ or }100$. The sign of each of the $\Delta_a$ is randomly selected to be positive or negative with equal probability. This is summarized in Table \ref{tab:unificationThresholdPrior}.
\begin{table}
\center
\begin{tabular}{|p{1.5in}|p{1.4in}|p{1.5in}|}
\hline
parameter & min & max\\
\hline
\hline
$M_u$ & $M_Z$ & $M_P$\\
\hline
$M_{\susy_c}$ & $M_Z$ & $M_P$\\
\hline
$h$ & 1 & 10, 100, or 1000\\
\hline
$|\Delta_a|$ & 1 & 10 or 100\\
\hline
$\alpha_u$ & 1/137 & 1\\
\hline
\end{tabular}
\caption{\small Minimal and maximal values of parameters used to compute the Bayesian evidence for the MSSM GUT with threshold corrections at the unification scale.}
\label{tab:unificationThresholdPrior}
\end{table}
Our results for the Bayesian evidences are given in Table \ref{tab:evidencesUnificationThreshold}.
\begin{table}
\center
\begin{tabular}{|p{1.5in}|p{1.4in}|p{1.5in}|}
\hline
 & $\Delta_{\max}=10$ & $\Delta_{\max}=100$\\
\hline
\hline
$h_{\max} = 10$ & $\mathcal Z = 4127\pm35$ & $\mathcal Z = 2092\pm25$ \\
\hline
$h_{\max} = 100$ & $\mathcal Z = 4536\pm37$ & $\mathcal Z = 2189\pm26$ \\
\hline
$h_{\max} = 1000$ & $\mathcal Z = 4673\pm38$ & $\mathcal Z = 2273\pm26$ \\
\hline
\end{tabular}
\caption{\small Bayesian evidences for the MSSM GUT with unification threshold corrections.}
\label{tab:evidencesUnificationThreshold}
\end{table}
The results show that the Bayesian evidence is not much effected by the introduction of the unification threshold corrections with $\Delta_{\max} = 10$. With $\Delta_{\max} = 100$, however, the Bayesian evidence is reduced by about a factor of two. Given that the results are so dependent upon the prior for the unification threshold corrections, we should be cautious about how we interpret the results.

\subsection{The $B-L$ MSSM GUT}
\label{sec:blmssmgut}
The $B-L$ MSSM GUT was proposed and studied in a series of papers\cite{Ovrut:2012wg,Marshall:2014kea,Marshall:2014cwa,Ovrut:2014rba,Ovrut:2015uea}. It involves a two-step breaking of the unified gauge theory by two Wilson lines, which are denoted $\chi_{T_{3R}}$ and $\chi_{B-L}$. The two Wilson lines each have a mass scale associated with them, denoted $M_{\chi_{T_{3R}}}$ and $M_{\chi_{B-L}}$, at which they partially break the gauge group. These two mass scales are not necessarily equal, and the gauge group in the intermediate regime between those mass scales depends on which mass scale is higher.

In the case that $\chi_{B-L}$ has the higher mass scale, it breaks the unified $SO(10)$ gauge group to a left-right type $SU(3)_C\times SU(2)_L\times SU(2)_R\times U(1)_{B-L}$ gauge group with slope factors,
\begin{eqnarray}
b_3^\lr=10 \quad b_2^\lr=14 \quad b_R^\lr=14 \quad b_{B-L}^\lr=19\ .
\label{eq:lrSlopeFactors}
\end{eqnarray}
At the lower scale, $M_{\chi_{T_{3R}}}$, the $\chi_{T_{3R}}$ Wilson line further breaks the $SU(2)_R$ factor to $U(1)_R$, yielding the $B-L$ MSSM gauge group, $SU(3)_C\times SU(2)_L\times U(1)_R\times U(1)_{B-L}$. We refer to this case as left-right type unification.

In the other case, that $\chi_{T_{3R}}$ has the higher mass scale, it breaks the unified $SO(10)$ gauge group to a Pati-Salam type $SU(4)_C\times SU(2)_L\times U(1)_R$ gauge group with slope factors,
\begin{eqnarray}
b_4^\ps=6 \quad b_2^\ps=14 \quad b_R^\ps=20\ .
\label{eq:psSlopeFactors}
\end{eqnarray}
At the lower scale, $M_{\chi_{B-L}}$, the $\chi_{B-L}$ Wilson line breaks the $SU(4)_C$ factor to $SU(3)_C\times U(1)_{B-L}$, yielding the $B-L$ MSSM gauge group, $SU(3)_C\times SU(2)_L\times U(1)_R\times U(1)_{B-L}$. We refer to this case as Pati-Salam type unification.

For this paper, the slope factors are all that is needed, but the complete particle content of the theory in between the two Wilson line scales is given in Ref.~\citen{Ovrut:2012wg}. Interestingly, in either the left-right type unification or the Pati-Salam type unification, the effect of this intermediate regime is to push the gauge couplings in the right direction to help them unify. This fact will result in a higher Bayesian evidence for the $B-L$ MSSM GUT.

\begin{figure}
\center
\includegraphics{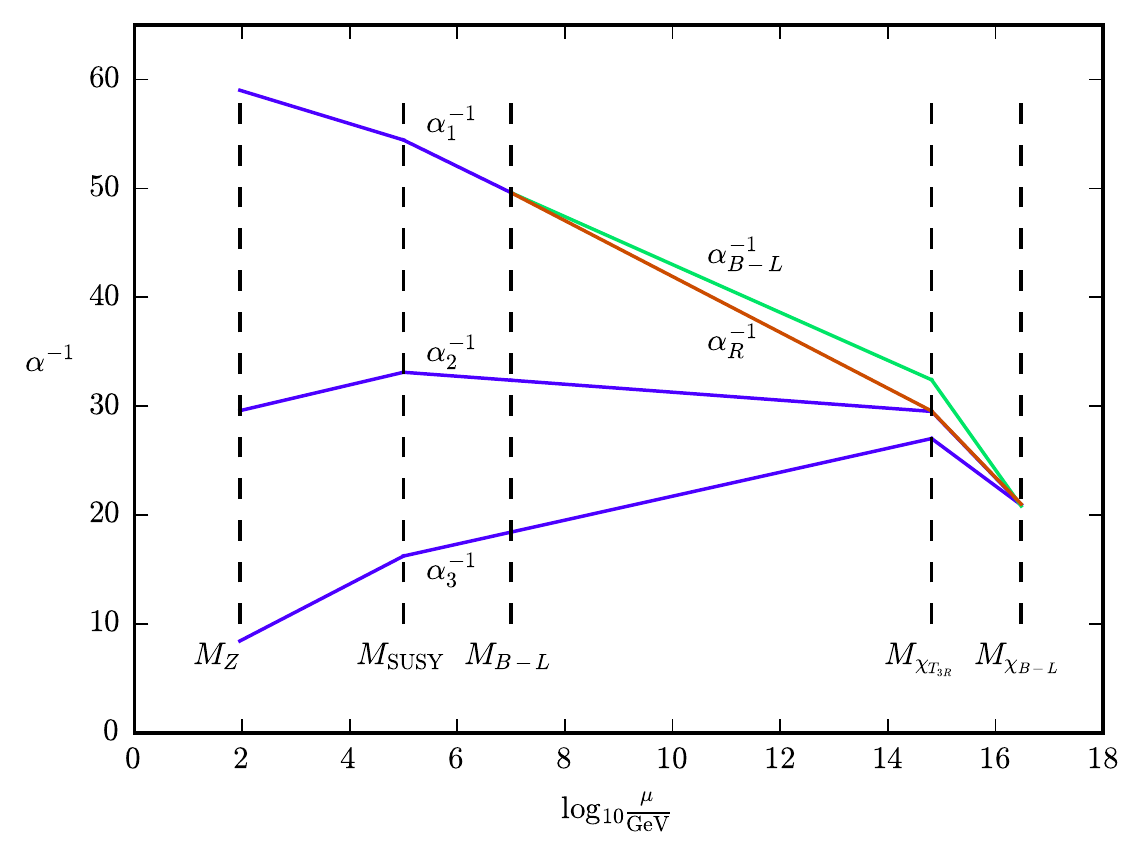}
\caption{\small Gauge coupling unification via the left-right type gauge group in the $B-L$ MSSM. Renormalization scale is denoted $\mu$. This plot was created using the values $M_\susy=10^5\mbox{ GeV}$, $M_{B-L}=10^7\mbox{ GeV}$, $\sin^2\theta_R=0.6$, $M_{\chi_{T_{3R}}}=6.5\times10^{14}\mbox{ GeV}$, and $M_{\chi_{B-L}}=3.0\times10^{16}$, along with the experimental values and slope factors given in the text. The large separation between the SUSY and $B-L$ scales, and between the intermediate and unification scales, may be unlikely, but these values are chosen for this plot to make their separation and the kinks in the plot more visually apparent. The values of the $\alpha_{B-L}$ and $\alpha_R$ exactly matching the value of $\alpha_1$ at the B-L scale is a coincidental result of the values used for this plot. It carries no physical significance. The general boundary condition is given in equation \eqref{eq:boundaryCondition}.}
\label{fig:blmssmGut}
\end{figure}

Below the scales of the two Wilson lines, the particle content of the theory is that of the MSSM plus three right-handed neutrino supermultiplets. The third-family right-handed sneutrino soft mass squared becomes negative due to one-loop radiative corrections, triggering radiative breaking of the $U(1)_R\times U(1)_{B-L}$ factor to $U(1)_Y$, yielding the familiar MSSM gauge group, $SU(3)_C\times SU(2)_L\times U(1)_Y$. This symmetry breaking is analogous to radiative electroweak symmetry breaking in the MSSM. The associated scale is called the $B-L$ scale, $M_{B-L}$. The boundary condition relating the $U(1)$ gauge couplings at the $B-L$ scale is
\begin{eqnarray}
\alpha_1^{-1}=\frac{2}{5}\alpha_{B-L}^{-1}+\frac35\alpha_{R}^{-1}\ .
\label{eq:boundaryCondition}
\end{eqnarray}
The parameter $\sin^2\theta_R$ is defined in a way analogous to the Weinberg angle,
\begin{eqnarray}
\sin^2\theta_R=\frac{\alpha_{B-L}^{-1}}{\frac23\alpha_R^{-1}+\alpha_{B-L}^{-1}}\ ,
\end{eqnarray}
where the gauge couplings are evaluated at the $B-L$ scale.
Finally, at the SUSY scale, $M_\susy$, the superpartners are integrated out and the standard model is obtained. We will consider the $B-L$ MSSM GUT both with and without the SUSY scale being split into $M_{\susy_c}$ and $M_{\susy_n}$. For more details on the $B-L$ MSSM see Refs.~\citen{Ovrut:2012wg,Ovrut:2015uea}. An example unification scenario is shown in Fig. \ref{fig:blmssmGut}. It was shown in Refs.~\citen{Ovrut:2012wg,Ovrut:2015uea} that the exact value of $M_{B-L}$ has no impact on the unification of gauge couplings, so for the present analysis it can be ignored. 

Applying the boundary conditions and slope factors for left-right type unification, the relationship between the observables ($\alpha_3$, $\alpha_2$, $\alpha_1$) and the parameters ($M_\susy$, $M_{\chi_{T_{3R}}}$, $M_{B-L}$, $\alpha_u$) is
\begin{eqnarray}
\alpha_{3}^{-1}&=&\alpha_u^{-1}+\frac{b_{3}^\lr}{2\pi}\ln\frac{M_{\chi_{B-L}}}{M_{\chi_{T_{3R}}}}+\frac{b_{3}^\mssm}{2\pi}\ln\frac{M_{\chi_{T_{3R}}}}{M_\susy}+\frac{b_{3}^\sm}{2\pi}\ln\frac{M_\susy}{M_Z}\nonumber\\
\alpha_{2}^{-1}&=&\alpha_u^{-1}+\frac{b_{2}^\lr}{2\pi}\ln\frac{M_{\chi_{B-L}}}{M_{\chi_{T_{3R}}}}+\frac{b_{2}^\mssm}{2\pi}\ln\frac{M_{\chi_{T_{3R}}}}{M_\susy}+\frac{b_{2}^\sm}{2\pi}\ln\frac{M_\susy}{M_Z}\nonumber\\
\alpha_{1}^{-1}&=&\alpha_u^{-1}+\frac{\frac25 b_{B-L}^\lr+\frac35 b_R^\lr}{2\pi}\ln\frac{M_{\chi_{B-L}}}{M_{\chi_{T_{3R}}}}\nonumber\\
&&+\frac{b_{1}^\mssm}{2\pi}\ln\frac{M_{\chi_{T_{3R}}}}{M_\susy}+\frac{b_{1}^\sm}{2\pi}\ln\frac{M_\susy}{M_Z}\ .
\label{eq:blmssmLrRelationship}
\end{eqnarray}
For Pati-Salam type unification, the relationship is 
\begin{eqnarray}
\alpha_{3}^{-1}&=&\alpha_u^{-1}+\frac{b_{3}^\ps}{2\pi}\ln\frac{M_{\chi_{T_{3R}}}}{M_{\chi_{B-L}}}+\frac{b_{3}^\mssm}{2\pi}\ln\frac{M_{\chi_{B-L}}}{M_\susy}+\frac{b_{3}^\sm}{2\pi}\ln\frac{M_\susy}{M_Z}\nonumber\\
\alpha_{2}^{-1}&=&\alpha_u^{-1}+\frac{b_{2}^\ps}{2\pi}\ln\frac{M_{\chi_{T_{3R}}}}{M_{\chi_{B-L}}}+\frac{b_{2}^\mssm}{2\pi}\ln\frac{M_{\chi_{B-L}}}{M_\susy}+\frac{b_{2}^\sm}{2\pi}\ln\frac{M_\susy}{M_Z}\nonumber\\
\alpha_{1}^{-1}&=&\alpha_u^{-1}+\frac{\frac25 b_{B-L}^\ps+\frac35 b_R^\ps}{2\pi}\ln\frac{M_{\chi_{T_{3R}}}}{M_{\chi_{B-L}}}\nonumber\\
&&+\frac{b_{1}^\mssm}{2\pi}\ln\frac{M_{\chi_{B-L}}}{M_\susy}+\frac{b_{1}^\sm}{2\pi}\ln\frac{M_\susy}{M_Z}\ .
\label{eq:blmssmPsRelationship}
\end{eqnarray}

Regarding the prior for the scales $M_{\chi_{B-L}}$ and $M_{\chi_{T_{3R}}}$, we take an approach similar to the one we use for the SUSY threshold. We consider both an unconstrained prior, where the scales may take any values between $M_Z$ and $M_P$, and a constrained one. The unconstrained prior is summarized in Table \ref{tab:unconstrainedBlmssmPrior}.
\begin{table}
\center
\begin{tabular}{|p{1.5in}|p{1.4in}|p{1.5in}|}
\hline
parameter & min & max\\
\hline
\hline
$M_{\chi_{B-L}}$ & $M_Z$ & $M_P$\\
\hline
$M_{\chi_{T_{3R}}}$ & $M_Z$ & $M_P$\\
\hline
$M_{\susy}$ & $M_Z$ & $M_P$\\
\hline
$\alpha_u$ & 1/137 & 1\\
\hline
\end{tabular}
\caption{\small Minimal and maximal values of parameters used to compute the Bayesian evidence for the $B-L$ MSSM GUT with unconstrained prior.}
\label{tab:unconstrainedBlmssmPrior}
\end{table}
For the constrained prior, we introduce a parameter $f$ defined by
\begin{eqnarray}
M_{\chi_{B-L}}=fM_{\chi_{T_{3R}}}\ ,
\end{eqnarray}
and assign to it a log-uniform prior with maximal and minimal values. Then $M_{\chi_{T_{3R}}}$ is no longer treated as a fundamental parameter, it is calculated from $M_{\chi_{B-L}}$ and $f$, which are treated as fundamental parameters. We consider $f_{\max}$ values of 10, 100, and 1000, with $f_{\min}$ taking values of 1/10, 1/100, or 1/1000 respectively. Note that, unlike the case of the SUSY threshold where $M_{\susy_c}$ was always the higher scale due to theoretical considerations, we are allowing either $M_{\chi_{T_{3R}}}$ or $M_{\chi_{B-L}}$ to be higher, and thus allowing either left-right type or a Pati-Salam type unification. The constrained prior is summarized in Table \ref{tab:constrainedBlmssmPrior}.
\begin{table}
\center
\begin{tabular}{|p{1.5in}|p{1.4in}|p{1.5in}|}
\hline
parameter & min & max\\
\hline
\hline
$M_{\chi_{B-L}}$ & $M_Z$ & $M_P$\\
\hline
$f$ & 1/10, 1/100, or 1/1000 & 10, 100, or 1000 (respectively)\\
\hline
$M_{\susy}$ & $M_Z$ & $M_P$\\
\hline
$\alpha_u$ & 1/137 & 1\\
\hline
\end{tabular}
\caption{\small Minimal and maximal values of parameters used to compute the Bayesian evidence for the $B-L$ MSSM GUT with the constrained prior.}
\label{tab:constrainedBlmssmPrior}
\end{table}
Step functions are used in the likelihood to enforce $M_Z \leq M_{\susy_n} \leq M_{\susy_c} \leq \min(M_{\chi_{T_{3R}}},M_{\chi_{B-L}})$. The Bayesian evidences with the constrained and unconstrained priors are given in Table \ref{tab:evidencesBlmssm}.
\begin{table}
\center
\begin{tabular}{|p{3.4in}|p{1.1in}|}
\hline
model & Bayesian evidence, $\mathcal{Z}$\\
\hline
\hline
$B-L$ MSSM GUT unconstrained prior & $2990\pm30$\\
\hline
$B-L$ MSSM GUT constrained prior, $f_{\min}=1/10$, $f_{\max}=10$ & $5635\pm41$\\
\hline
$B-L$ MSSM GUT constrained prior, $f_{\min}=1/100$, $f_{\max}=100$ & $5755\pm42$\\
\hline
$B-L$ MSSM GUT constrained prior, $f_{\min}=1/1000$, $f_{\max}=1000$ & $5052\pm39$\\
\hline
\end{tabular}
\caption{\small Bayesian evidences for the $B-L$ MSSM GUT with unconstrained and constrained priors.}
\label{tab:evidencesBlmssm}
\end{table}
One thing to note when comparing these results to the results for the MSSM GUT with SUSY threshold corrections in Table \ref{tab:susyThresholdGutEvidences}, is that the $B-L$ MSSM GUT is favored over the MSSM GUT with SUSY threshold corrections when the unconstrained prior is used for both. This is due to the fact that in the $B-L$ MSSM GUT can be successful with either left-right type or Pati-Salam type unification, that is, with either $M_{\chi_{T_{3R}}} < M_{\chi_{B-L}}$ or $M_{\chi_{B-L}} < M_{\chi_{T_{3R}}}$. The MSSM GUT with SUSY threshold corrections, in contrast, only successfully unifies with $M_{\susy_n} < M_{\susy_c}$. When constrained priors are used, neither the $B-L$ MSSM GUT nor the MSSM GUT with SUSY threshold corrections is substantially favored over the other. This is due to the fact that with the constrained prior we always have $M_{\susy_n} < M_{\susy_c}$ so the fact that the MSSM GUT does not not unify otherwise is moot. Recall that the constrained prior was chosen this way because theoretical considerations suggested it.

With the constrained prior, the Bayesian evidence does not depend strongly on the values of $f_{\min}$ and $f_{\max}$. We nevertheless regard $f_{\min}=1/10$, $f_{\max}=10$ as the most reasonable choice because each Wilson line scale is related to the inverse radius of a non-contractible curve in the Calabi-Yau manifold of the underlying string theory\cite{Ovrut:2012wg}. They are thus both related to the string compactification scale and there is no reason to suspect that they should be different by more than an order of magnitude. See Ref.~\citen{Ovrut:2012wg} for discussion.

\begin{figure}
\center
\includegraphics{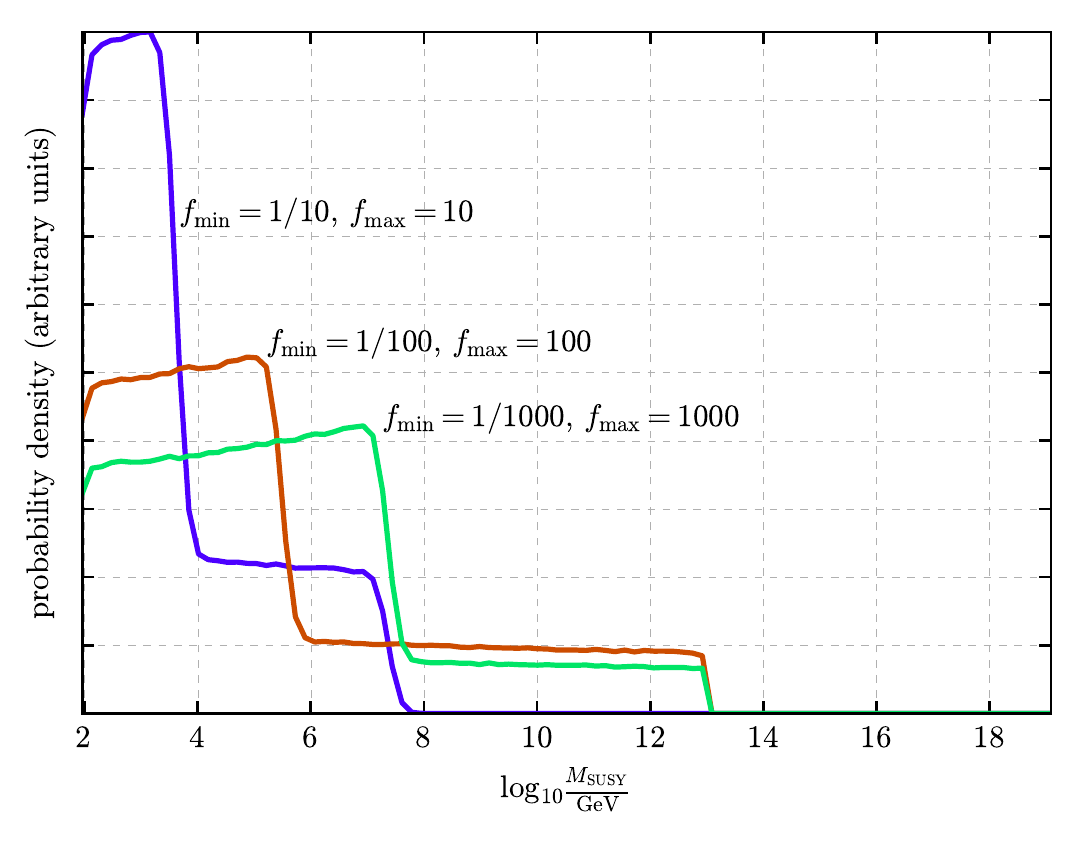}
\caption{\small Posterior probability distribution for $\log M_{\susy}$ in arbitrary units. The small fluctuations in the lines are due to the statistical uncertainty in the Monte Carlo integration. The three curves are normalized relative to each other so that they all have the same area underneath.}
\label{fig:posteriorBlmssm}
\end{figure}

Fig. \ref{fig:posteriorBlmssm} shows the posterior probability distribution of $\log M_{\susy}$ in $B-L$ MSSM GUT with constrained priors. The result is plotted in arbitrary units. This figure is limited to the constrained priors because they are better motivated and favored by the Bayesian model comparison. The figure shows that the SUSY scale is not constrained to be weak-scale. It can be orders of magnitude higher and still be consistent with unification. However, exactly how much higher the SUSY scale can be depends on the prior chosen. With $f_{\min}=1/10$, $f_{\max} = 10$ it can be up to around $10^7$ GeV. With $f_{\min}=1/1000$, $f_{\max} = 1000$, it can be as high as $10^{13}$ GeV.

The ``two-step'' shape of the posteriors is due to the two different unification schemes, left-right type and Pati-Salam type. Comparing the left-right type slope factors in equation \eqref{eq:lrSlopeFactors} to the Pati-Salam type slope factors in equation \eqref{eq:psSlopeFactors}, we see that the latter slope factors are more different from each other, allowing a stronger push toward unification with less separation of the Wilson line scales. This means the Pati-Salam type unification scheme can accommodate a higher SUSY scale and still achieve unification, resulting in the longer, lower ``step'' in each of the three posterior distributions shown in Fig. \ref{fig:posteriorBlmssm}. The abrupt cutoff at $M_{\susy}\approx 10^{13}$ GeV is due to the step function assigning zero likelihood to points for which $M_{\susy} > M_{\chi_{B-L}}$.

The $B-L$ MSSM GUT may of course include the same threshold corrections that we included in the MSSM GUT. That is, SUSY threshold corrections modeled by separating the SUSY scale into $M_{\susy_c}$ and $M_{\susy_n}$, and unification threshold corrections modeled by the parameters $\Delta_1$, $\Delta_2$, and $\Delta_3$. Considering first only the SUSY threshold corrections, the relationship between the parameters and observables is equations \eqref{eq:blmssmLrRelationship} and \eqref{eq:blmssmPsRelationship} appropriately modified to include the SUSY threshold terms in equation \eqref{eq:mssmSusyThresholdRelationship}. That is, in the case of left-right type unification, 
\begin{eqnarray}
\alpha_{3}^{-1}&=&\alpha_u^{-1}+\frac{b_{3}^\lr}{2\pi}\ln\frac{M_{\chi_{B-L}}}{M_{\chi_{T_{3R}}}}+\frac{b_{3}^\mssm}{2\pi}\ln\frac{M_{\chi_{T_{3R}}}}{M_{\susy_c}}\nonumber\\
&&+\frac{b_{3}^\ncmssm}{2\pi}\ln\frac{M_{\susy_c}}{M_{\susy_n}}+\frac{b_{3}^\sm}{2\pi}\ln\frac{M_{\susy_n}}{M_Z}\nonumber\\
\alpha_{2}^{-1}&=&\alpha_u^{-1}+\frac{b_{2}^\lr}{2\pi}\ln\frac{M_{\chi_{B-L}}}{M_{\chi_{T_{3R}}}}+\frac{b_{2}^\mssm}{2\pi}\ln\frac{M_{\chi_{T_{3R}}}}{M_{\susy_c}}\nonumber\\
&&+\frac{b_{2}^\ncmssm}{2\pi}\ln\frac{M_{\susy_c}}{M_{\susy_n}}+\frac{b_{2}^\sm}{2\pi}\ln\frac{M_{\susy_n}}{M_Z}\nonumber\\
\alpha_{1}^{-1}&=&\alpha_u^{-1}+\frac{\frac25 b_{B-L}^\lr+\frac35 b_R^\lr}{2\pi}\ln\frac{M_{\chi_{B-L}}}{M_{\chi_{T_{3R}}}}+\frac{b_{1}^\mssm}{2\pi}\ln\frac{M_{\chi_{T_{3R}}}}{M_{\susy_c}}\nonumber\\
&&+\frac{b_{1}^\ncmssm}{2\pi}\ln\frac{M_{\susy_c}}{M_{\susy_n}}+\frac{b_{1}^\sm}{2\pi}\ln\frac{M_{\susy_n}}{M_Z}\ .
\label{eq:blmssmLrSusyThresholdRelationship}
\end{eqnarray}
And the case of Pati-Salam type unification,
\begin{eqnarray}
\alpha_{3}^{-1}&=&\alpha_u^{-1}+\frac{b_{3}^\ps}{2\pi}\ln\frac{M_{\chi_{T_{3R}}}}{M_{\chi_{B-L}}}+\frac{b_{3}^\mssm}{2\pi}\ln\frac{M_{\chi_{T_{3R}}}}{M_{\susy_c}}\nonumber\\
&&+\frac{b_{3}^\ncmssm}{2\pi}\ln\frac{M_{\susy_c}}{M_{\susy_n}}+\frac{b_{3}^\sm}{2\pi}\ln\frac{M_{\susy_n}}{M_Z}\nonumber\\
\alpha_{2}^{-1}&=&\alpha_u^{-1}+\frac{b_{2}^\ps}{2\pi}\ln\frac{M_{\chi_{T_{3R}}}}{M_{\chi_{B-L}}}+\frac{b_{2}^\mssm}{2\pi}\ln\frac{M_{\chi_{T_{3R}}}}{M_{\susy_c}}\nonumber\\
&&+\frac{b_{2}^\ncmssm}{2\pi}\ln\frac{M_{\susy_c}}{M_{\susy_n}}+\frac{b_{2}^\sm}{2\pi}\ln\frac{M_{\susy_n}}{M_Z}\nonumber\\
\alpha_{1}^{-1}&=&\alpha_u^{-1}+\frac{\frac25 b_{B-L}^\ps+\frac35 b_R^\ps}{2\pi}\ln\frac{M_{\chi_{T_{3R}}}}{M_{\chi_{B-L}}}+\frac{b_{1}^\mssm}{2\pi}\ln\frac{M_{\chi_{T_{3R}}}}{M_{\susy_c}}\nonumber\\
&&+\frac{b_{1}^\ncmssm}{2\pi}\ln\frac{M_{\susy_c}}{M_{\susy_n}}+\frac{b_{1}^\sm}{2\pi}\ln\frac{M_{\susy_n}}{M_Z}\ .
\label{eq:blmssmPsSusyThresholdRelationship}
\end{eqnarray}
We use a constrained prior with $h_{\max}=10$ and $f_{\min}=1/10$, $f_{\max}=10$. This is summarized in Table \ref{tab:priorBlmssmSusyThresholdGut}.
\begin{table}
\center
\begin{tabular}{|p{1.5in}|p{1.4in}|p{1.5in}|}
\hline
parameter & min & max\\
\hline
\hline
$M_{\chi_{B-L}}$ & $M_Z$ & $M_P$\\
\hline
$f$ & 1/10 & 10\\
\hline
$M_{\susy_c}$ & $M_Z$ & $M_P$\\
\hline
$h$ & 1 & 10\\
\hline
$\alpha_u$ & 1/137 & 1\\
\hline
\end{tabular}
\caption{\small Minimal and maximal values of parameters used to compute the Bayesian evidence for the $B-L$ MSSM GUT with SUSY threshold corrections.}
\label{tab:priorBlmssmSusyThresholdGut}
\end{table}
Considering both the SUSY threshold corrections and the unification threshold corrections, the relationship between the parameters and observables is the same as in equations \eqref{eq:blmssmLrSusyThresholdRelationship} and \eqref{eq:blmssmPsSusyThresholdRelationship} with the $\Delta_1/(4\pi)$, $\Delta_2/(4\pi)$, and $\Delta_3/(4\pi)$ added to the three equations respectively, as in equation \eqref{eq:mssmSusyThresholdNuisanceRelationship}. We use the same prior as in the $B-L$ MSSM GUT with SUSY threshold corrections but with unification threshold corrections prior with $\Delta_{\max}=10$. This choice is made because $\Delta_{\max}=10$ had a substantially higher Bayesian evidence than $\Delta_{\max}=100$ in the case of the MSSM GUT with unification threshold corrections. This is summarized in Table \ref{tab:priorBlmssmUnificationThresholdGut}. As in the case of the MSSM GUT with unification threshold corrections, the sign of each $\Delta_a$ is randomly selected to be positive or negative with equal probability.
\begin{table}
\center
\begin{tabular}{|p{1.5in}|p{1.4in}|p{1.5in}|}
\hline
parameter & min & max\\
\hline
\hline
$M_{\chi_{B-L}}$ & $M_Z$ & $M_P$\\
\hline
$f$ & 1/10 & 10\\
\hline
$M_{\susy_c}$ & $M_Z$ & $M_P$\\
\hline
$h$ & 1 & 10\\
\hline
$|\Delta_a|$ & 1 & 10\\
\hline
$\alpha_u$ & 1/137 & 1\\
\hline
\end{tabular}
\caption{\small Minimal and maximal values of parameters used to compute the Bayesian evidence for the $B-L$ MSSM GUT with SUSY threshold corrections and unification threshold corrections.}
\label{tab:priorBlmssmUnificationThresholdGut}
\end{table}

The Bayesian evidences are given in Table \ref{tab:evidencesBlmssmThresholdGut}.
\begin{table}
\center
\begin{tabular}{|p{3.4in}|p{1.1in}|}
\hline
model & Bayesian evidence, $\mathcal{Z}$\\
\hline
\hline
$B-L$ MSSM GUT with SUSY threshold & $5655\pm41$\\
\hline
$B-L$ MSSM GUT with SUSY threshold and unification threshold & $4877\pm38$\\
\hline
\end{tabular}
\caption{\small Bayesian evidences for the $B-L$ MSSM GUT with SUSY threshold corrections and with both SUSY threshold corrections and unification threshold corrections.}
\label{tab:evidencesBlmssmThresholdGut}
\end{table}
Note that the word ``threshold'' is used for brevity in the tables instead of ``threshold corrections''. The results show that including the threshold corrections in the $B-L$ MSSM GUT does not substantially change the Bayesian evidence.

\subsection{A non-SUSY GUT}

With sufficiently large unification threshold corrections, there arises the possibility of unification without any SUSY below the unification scale. The tools developed in this paper allow us to consider such a non-SUSY GUT easily, if we assume that the slope factors are (at least approximately) those of the SM. The relationship between the parameters and observables is simply
\begin{eqnarray}
\alpha_a^{-1}=\alpha_u^{-1}+\frac{\Delta_a}{4\pi}+\frac{b_a^\sm}{2\pi}\ln\frac{M_u}{M_Z}\ ,
\end{eqnarray}
where $a\in {1,2,3}$. For a prior we consider $\Delta_{\max} = 100 \mbox{ or } 1000$. Using $\Delta_{\max}=10$ does not yield large enough threshold corrections to permit unification in this model (that is, the Bayesian evidence would be zero). The prior is summarized in Table \ref{tab:smPriors}.
\begin{table}
\center
\begin{tabular}{|p{1.5in}|p{1.4in}|p{1.5in}|}
\hline
parameter & min & max\\
\hline
\hline
$M_u$ & $M_Z$ & $M_P$\\
\hline
$|\Delta_a|$ & 1 & 100, or 1000\\
\hline
$\alpha_u$ & 1/137 & 1\\
\hline
\end{tabular}
\caption{\small Minimal and maximal values of parameters used to compute the Bayesian evidence for the non-SUSY GUT.}
\label{tab:smPriors}
\end{table}
The results are given in Table \ref{tab:evidencesSM}.
\begin{table}
\center
\begin{tabular}{|p{3.4in}|p{1.1in}|}
\hline
model & Bayesian evidence, $\mathcal{Z}$\\
\hline
\hline
non-SUSY GUT, $\Delta_{\max}=100$ & $728\pm15$\\
\hline
non-SUSY GUT, $\Delta_{\max}=1000$ & $323\pm10$\\
\hline
\end{tabular}
\caption{\small Bayesian evidences for the $B-L$ MSSM GUT with unconstrained and constrained priors.}
\label{tab:evidencesSM}
\end{table}
The results demonstrate that even though large threshold corrections may allow unification without SUSY, the SUSY GUTs are still favored by the Bayesian analysis.

\section{Conclusion and Discussion}
\label{sec:4}

This paper gives 22 Bayesian evidences, all of which are repeated in Table \ref{tab:evidences}.
\begin{table}
\center
\begin{tabular}{|p{3.4in}|p{1.1in}|}
\hline
model & Bayesian evidence, $\mathcal{Z}$\\
\hline
\hline
puzzle model & 124\\
\hline
\hline
weak-scale MSSM GUT & $(1.0569\pm.0014)\times 10^5$\\
\hline
\hline
MSSM GUT with SUSY threshold & \\
\hline
\quad unconstrained prior & $1204\pm19$\\
\hline
\quad $h_{\max}=10$ & $5316\pm40$ \\
\hline
\quad $h_{\max}=100$ & $5189\pm40$ \\
\hline
\quad $h_{\max}=1000$ & $5058\pm39$ \\
\hline
\hline
MSSM GUT with SUSY threshold and unification threshold & \\
\hline
\quad $h_{\max}=10$, $\Delta_{\max}=10$ & $4127\pm35$ \\
\hline
\quad $h_{\max}=100$, $\Delta_{\max}=10$ & $4536\pm37$ \\
\hline
\quad $h_{\max}=1000$, $\Delta_{\max}=10$ & $4673\pm38$ \\
\hline
\quad $h_{\max}=10$, $\Delta_{\max}=100$ & $2092\pm25$ \\
\hline
\quad $h_{\max}=100$, $\Delta_{\max}=100$ & $2189\pm26$ \\
\hline
\quad $h_{\max}=1000$, $\Delta_{\max}=100$ & $2273\pm26$ \\
\hline
\hline
$B-L$ MSSM GUT & \\
\hline
\quad unconstrained prior & $2990\pm30$\\
\hline
\quad $f_{\min}=1/10$, $f_{\max}=10$ & $5635\pm41$\\
\hline
\quad $f_{\min}=1/100$, $f_{\max}=100$ & $5755\pm42$\\
\hline
\quad $f_{\min}=1/1000$, $f_{\max}=1000$ & $5052\pm39$\\
\hline
\hline
$B-L$ MSSM GUT with SUSY threshold & \\
\hline
\quad $h_{\max}=10$, $f_{\min} = 1/10$, $f_{\max} = 10$ & $5655\pm41$\\
\hline
\hline
$B-L$ MSSM GUT with SUSY threshold and unification threshold& \\
\hline
\quad $h_{\max}=10$, $f_{\min} = 1/10$, $f_{\max} = 10$, $\Delta_{\max} = 10$ & $4877\pm38$\\
\hline
\hline
non-SUSY GUT & \\
\hline
\quad $\Delta_{\max}=100$ & $728\pm15$\\
\hline
\quad $\Delta_{\max}=1000$ & $323\pm10$\\
\hline
\end{tabular}
\caption{\small All Bayesian evidences given in this paper.}
\label{tab:evidences}
\end{table}
From these and our other results we make the following conclusions.
\begin{itemize}
\item{In our results, every GUT is substantially more supported than the puzzle (non-unifying) model. Even GUTs that have more parameters than the puzzle model are more supported. This implies that even though the additional parameters may be in conflict with simplicity, they more than make up for it by enabling unification. This is due to the fact that the observables are relatively insensitive to the additional parameters in the GUTs, corresponding to a degree of naturalness. An intuitive understanding of simplicity and naturalness would not be sufficient to reach this conclusion because it offers no way to weigh naturalness against the lack of simplicity. A quantitative language that can simultaneously weigh both simplicity and naturalness is needed and Bayesian model comparison provides just that.}
\item{The constrained priors for the SUSY threshold corrections and Wilson line scales are more supported than the unconstrained priors. This is reassuring because the constrained priors are better motivated theoretically.}
\item{In the MSSM GUT and the $B-L$ MSSM GUT, adding unification threshold corrections slightly lowers the Bayesian evidence. While there may be other reasons for believing in significant unification threshold corrections, from the standpoint of gauge unification in these SUSY GUTs, large unification threshold corrections are unnecessary.}
\item{The most strongly supported model (with the exception of the weak-scale MSSM GUT, which has $M_{\susy}\approx M_Z$, incompatible with LHC data) is the $B-L$ MSSM GUT. That said, its support over the MSSM GUT, while statistically significant with regard to the Monte Carlo integration, is so slight as to be barely worth mentioning. We still regard the $B-L$ MSSM as the stronger model for reasons other than these results (it provides a theory of $R$-parity and neutrino masses, and is motivated by string theory).}
\item{The SUSY GUTs are all more strongly supported than the non-SUSY GUT that relies on large unification threshold corrections. While it may be tempting to rely on large unification threshold corrections to allow unification in models which do not otherwise allow it, the SUSY models, which do not rely on such large unification threshold corrections, are more strongly supported.}
\item{Based on posteriors for the SUSY scale in Figs. \ref{fig:posteriorSusyThreshold} and \ref{fig:posteriorBlmssm}, SUSY GUTs are consistent with SUSY scales well above current LHC bounds. It would be a mistake to think that unification is a feature unique to weak scale SUSY. That said, the posteriors do tend to be weighted more toward lower-SUSY scales, and this speaks to the utility of the Very Large Hadron Collider (VLHC) or other next generation collider (see Ref.~\citen{Fowlie:2014xha} for a paper on this topic).}
\item{We showed that the quantified Bayesian naturalness, unlike the BG sensitivity, suggests that the proton mass is natural, consistent with physicists' intuitive notions of naturalness. Bayesian naturalness thus accomplishes what had already been accomplished by some ad-hoc refinements of the BG sensitivity\cite{Anderson:1994dz,Athron:2007ry}. But Bayesian naturalness follows from principles that are more basic rather than ad-hoc.}
\end{itemize}

While Bayesian model comparison has already been extensively applied to supersymmetry, it has seen relatively little use in other areas of theoretical physics (Refs.~\citen{Clarke:2016jzm,Fowlie:2016jlx} and now this paper are three examples). We are led to wonder what other open questions in fundamental physics may be productively addressed using Bayesian model comparison. Where neither experimental data nor intuitive notions of naturalness, simplicity, and testability are able to decisively settle a question, Bayesian model comparison may be useful. String theory, eternal inflation, and the anthropic principle all come to mind. Some of these have been mentioned in Ref.~\citen{FowlieTalk}.

Lastly we should point out that quantitative analysis using Bayesian model comparison should not replace qualitative analysis using intuitive notions of naturalness, simplicity, and testability. On the contrary, because analysis using Bayesian model comparison can be computationally intensive, and often only verifies things that are obvious to the experienced physicist, Bayesian model comparison should take its place alongside intuitive notions of naturalness, simplicity, and testability as useful tools for guiding physics research. In fact Bayesian model comparison provides a strong justification for the continued of those intuitive notions.

\appendix

\section{Monte Carlo Integration on the GPU}
\label{sec:appendix}

The results in equations \eqref{eq:zpuzzle} and \eqref{eq:zgut} were both verified using Monte Carlo integration\footnote{See Ref.~\citen{recipes} for an explanation of Monte Carlo integration.} implemented in single-threaded Python code. Due to the larger number of parameters in the other GUTs considered, the Bayesian evidences do not have straightforward analytic solutions, so we rely entirely on Monte Carlo integration for those results. We determined that it would be necessary to implement the Monte Carlo integration using the parallel processing power of a Graphics Processing Unit (GPU) to get acceptable performance. We have released our code to the public\cite{git}.

Monte Carlo integration was chosen over other methods of numerical integration for three reasons. First, these are integrals over multiple variables. Many other methods of integration take time that is exponential in the number of variables of integration. Monte Carlo integration doesn't. Second, with Monte Carlo integration, the standard deviation of the samples can be used to straightforwardly obtain an estimate of the uncertainty in the result. Third, because each sample is independent of all the others, Monte Carlo integration lends itself easily to parallelization. Therefore it can better utilize the capabilities of modern computer hardware, including GPUs.

We implemented the Monte Carlo integration using OpenGL 4.0 and the C programming language, along with the Simple DirectMedia Layer 2.0 library for window and OpenGL context creation, and the Epoxy library for OpenGL function pointer management. The Monte Carlo integration was executed on a computer with an NVIDIA GTX 960 GPU, running Xubuntu 16.04 with NVIDIA proprietary graphics drivers. Creating a $1000\times500$ pixel window and drawing a single quad over the entire window results in the fragment shader being invoked at each of the $1000\times500$ pixels. The GPU then executes fragment shader invocations in parallel, as is the standard operation for a GPU in computer graphics applications. We implement the Monte Carlo sample generation in the fragment shaders using the OpenGL Shading Language (GLSL). This approach was used instead of compute shaders because early development was using OpenGL 3.3, which does not support compute shaders.

In each fragment shader invocation, we generate many Monte Carlo samples, average them, and output the average as a 32-bit floating point number in one of the color channels normally used for graphical output to the screen. Since this is not a graphics application, we instruct OpenGL to render to a texture in video memory, rather than to the screen. The C code then reads this texture data into main memory where the output from all the fragment shader invocations can be averaged and the final result computed.

OpenGL was chosen over a ready-made GPU compute application or library because, to our knowledge, existing ready-made GPU compute applications and libraries do not utilize state of the art Philox pseudo-random number generation (discussed below). OpenGL was chosen over CUDA for platform independence, and it was chosen over OpenCL because the authors are more familiar with it and it has all the needed functionality.

The uncertainty in the result of a Monte Carlo integration is proportional to one over the square root of the number of samples. Obtaining a fractional uncertainty below 1\% required $\mathcal O(10^{12})$ samples.\footnote{Such a large number of samples is required because the integrand contains extremely narrow Gaussian likelihood functions, which are nearly zero over the vast majority of the volume over which we integrate.} This presents a problem for popular pseudo-random number generators (PRNGs). The problem, and its solution, which have been thoroughly understood in Ref.~\citen{Manssen:2012vj,philox}, are briefly reviewed in this paragraph and the next. The problem with most popular PRNGs is that they require each pseudo-random number in a sequence of pseudo-random numbers to be generated successively, rather than in parallel. Generating $\mathcal O(10^{12})$ pseudo-random numbers successively on the CPU would take too long, and storing them would exhaust RAM. Generating multiple smaller sequences in parallel is a partial solution, but it is still RAM and CPU intensive and in practice it is difficult to guarantee that the smaller sequences are not correlated in a way that could lead to systematic error.

The solution to this problem is to instead use counter-based PRNGs, wherein each pseudo-random number, $x_n$ is simply defined by a function, $b$, applied to a counter, $n$:
\begin{eqnarray}
x_n = b(n)\ .
\end{eqnarray}
The entire sequence can be generated in parallel. We use the Philox counter-based PRNG introduced in Ref.~\citen{philox} and further tested in Ref.~\citen{Manssen:2012vj}, which is highly performant, uses little memory, and has passed stringent tests for unwanted correlations. Specifically, we use the Philox-4$\times$32-7 counter-based PRNG.

Our GPU-based Monte Carlo integration was used to verify equation \eqref{eq:zgut}, and showed a 5000$\times$ speed increase over the single-threaded Python code.\footnote{This remarkable speed increase may be partially due to the fact that we made no attempt to optimize the Python code. Nevertheless, an $\mathcal O(1000)\times$ speed increase is to be expected since the NVIDIA GTX 960 GPU has about 1000 cores.} Running the code to compute any one of the Bayesian evidences given in this paper takes about 3 minutes with $10^{12}$ samples.

\section*{Acknowledgments}
The authors are grateful to an anonymous reviewer for their thorough reading and detailed comments that helped improve and expand the manuscript. A. Purves thanks Burt A. Ovrut and Sogee Spinner for mentorship and continued support. He thanks Rehan Deen for helpful conversations. He thanks the faculty, staff, and administration of Manhattanville College for support, encouragement, and helpful conversations. He thanks his students for their penetrating questions. P. Fundira thanks Edward Schwartz and the faculty of the Department of Mathematics and Computer Science for their encouragement and support. He thanks Bothwell and Tsungirirayi Fundira for the opportunity to attend Manhattanville College. He thanks Farirai A. Fundira for his companionship over the last four years. This work was not supported by any specific grant.


\begin{thebibliography}{99}

\bibitem{Giudice:2008bi} 
  G.~F.~Giudice,
  ``Naturally Speaking: The Naturalness Criterion and Physics at the LHC,''
  In *Kane, Gordon (ed.), Pierce, Aaron (ed.): Perspectives on LHC physics* 155-178
  [arXiv:0801.2562 [hep-ph]].

\bibitem{Nelson}
  P. Nelson,
  ``Naturalness in Theoretical Physics,''
  Am.\ Sci.\ {\bf 73}, 60 (1985)

\bibitem{Grinbaum:2009sk} 
  A.~Grinbaum,
  ``Which fine-tuning arguments are fine?,''
  Found.\ Phys.\  {\bf 42}, 615 (2012)
  doi:10.1007/s10701-012-9629-9
  [arXiv:0903.4055 [physics.hist-ph]].

\bibitem{Guth:1980zm} 
  A.~H.~Guth,
  ``The Inflationary Universe: A Possible Solution to the Horizon and Flatness Problems,''
  Phys.\ Rev.\ D {\bf 23}, 347 (1981).
  doi:10.1103/PhysRevD.23.347

\bibitem{Khoury:2001wf} 
  J.~Khoury, B.~A.~Ovrut, P.~J.~Steinhardt and N.~Turok,
  ``The Ekpyrotic universe: Colliding branes and the origin of the hot big bang,''
  Phys.\ Rev.\ D {\bf 64}, 123522 (2001)
  doi:10.1103/PhysRevD.64.123522
  [hep-th/0103239].

\bibitem{Bousso:2007gp} 
  R.~Bousso,
  ``TASI Lectures on the Cosmological Constant,''
  Gen.\ Rel.\ Grav.\  {\bf 40}, 607 (2008)
  doi:10.1007/s10714-007-0557-5
  [arXiv:0708.4231 [hep-th]].

\bibitem{Martin:2012bt} 
  J.~Martin,
  ``Everything You Always Wanted To Know About The Cosmological Constant Problem (But Were Afraid To Ask),''
  Comptes Rendus Physique {\bf 13}, 566 (2012)
  doi:10.1016/j.crhy.2012.04.008
  [arXiv:1205.3365 [astro-ph.CO]].

\bibitem{Peccei:2006as} 
  R.~D.~Peccei,
  ``The Strong CP problem and axions,''
  Lect.\ Notes Phys.\  {\bf 741}, 3 (2008)
  doi:10.1007/978-3-540-73518-2\_1
  [hep-ph/0607268].

\bibitem{Martin:1997ns} 
  S.~P.~Martin,
  ``A Supersymmetry primer,''
  Adv.\ Ser.\ Direct.\ High Energy Phys.\  {\bf 21}, 1 (2010)
  [Adv.\ Ser.\ Direct.\ High Energy Phys.\  {\bf 18}, 1 (1998)]
  doi:10.1142/9789812839657\_0001, 10.1142/9789814307505\_0001
  [hep-ph/9709356].

\bibitem{Barbieri:2000gf} 
  R.~Barbieri and A.~Strumia,
  ``The `LEP paradox',''
  hep-ph/0007265.

\bibitem{Nobbenhuis:2006yf} 
  S.~Nobbenhuis,
  ``The Cosmological Constant Problem, an Inspiration for New Physics,''
  Ph.D. thesis, Utrecht U. (2006),
  gr-qc/0609011.

\bibitem{Kim:2008hd} 
  J.~E.~Kim and G.~Carosi,
  ``Axions and the Strong CP Problem,''
  Rev.\ Mod.\ Phys.\  {\bf 82}, 557 (2010)
  doi:10.1103/RevModPhys.82.557
  [arXiv:0807.3125 [hep-ph]].

\bibitem{Kim:1979if} 
  J.~E.~Kim,
  ``Weak Interaction Singlet and Strong CP Invariance,''
  Phys.\ Rev.\ Lett.\  {\bf 43}, 103 (1979).
  doi:10.1103/PhysRevLett.43.103

\bibitem{Shifman:1979if} 
  M.~A.~Shifman, A.~I.~Vainshtein and V.~I.~Zakharov,
  ``Can Confinement Ensure Natural CP Invariance of Strong Interactions?,''
  Nucl.\ Phys.\ B {\bf 166}, 493 (1980).
  doi:10.1016/0550-3213(80)90209-6

\bibitem{Dine:1981rt} 
  M.~Dine, W.~Fischler and M.~Srednicki,
  ``A Simple Solution to the Strong CP Problem with a Harmless Axion,''
  Phys.\ Lett.\ B {\bf 104}, 199 (1981).
  doi:10.1016/0370-2693(81)90590-6

\bibitem{Diaz-Cruz:2016pmm} 
  J.~L.~Daz-Cruz, W.~G.~Hollik and U.~J.~Saldaa-Salazar,
  ``Addressing the strong CP problem with quark mass ratios,''
  arXiv:1605.03860 [hep-ph].

\bibitem{Banerjee:2000qw} 
  H.~Banerjee, D.~Chatterjee and P.~Mitra,
  ``Is there still a strong CP problem?,''
  Phys.\ Lett.\ B {\bf 573}, 109 (2003)
  doi:10.1016/j.physletb.2003.08.058
  [hep-ph/0012284].

\bibitem{Blinov:2016kte} 
  N.~Blinov and A.~Hook,
  JHEP {\bf 1606}, 176 (2016)
  doi:10.1007/JHEP06(2016)176
  [arXiv:1605.03178 [hep-ph]].

\bibitem{Hook:2014cda} 
  A.~Hook,
  ``Anomalous solutions to the strong CP problem,''
  Phys.\ Rev.\ Lett.\  {\bf 114}, no. 14, 141801 (2015)
  doi:10.1103/PhysRevLett.114.141801
  [arXiv:1411.3325 [hep-ph]].

\bibitem{Swain:2010rr} 
  J.~Swain,
  ``Black Holes and the Strong CP Problem,''
  arXiv:1005.1097 [gr-qc].

\bibitem{Csaki:2005fc} 
  C.~Csaki, G.~Marandella, Y.~Shirman and A.~Strumia,
  ``The Super-little Higgs,''
  Phys.\ Rev.\ D {\bf 73}, 035006 (2006)
  doi:10.1103/PhysRevD.73.035006
  [hep-ph/0510294].

\bibitem{Bellazzini:2009ix} 
  B.~Bellazzini, C.~Csaki, A.~Delgado and A.~Weiler,
  ``SUSY without the Little Hierarchy,''
  Phys.\ Rev.\ D {\bf 79}, 095003 (2009)
  doi:10.1103/PhysRevD.79.095003
  [arXiv:0902.0015 [hep-ph]].

\bibitem{Babu:2008ge} 
  K.~S.~Babu, I.~Gogoladze, M.~U.~Rehman and Q.~Shafi,
  ``Higgs Boson Mass, Sparticle Spectrum and Little Hierarchy Problem in Extended MSSM,''
  Phys.\ Rev.\ D {\bf 78}, 055017 (2008)
  doi:10.1103/PhysRevD.78.055017
  [arXiv:0807.3055 [hep-ph]].

\bibitem{Dermisek:2005ar} 
  R.~Dermisek and J.~F.~Gunion,
  ``Escaping the large fine tuning and little hierarchy problems in the next to minimal supersymmetric model and $h \rightarrow aa$ decays,''
  Phys.\ Rev.\ Lett.\  {\bf 95}, 041801 (2005)
  doi:10.1103/PhysRevLett.95.041801
  [hep-ph/0502105].

\bibitem{Farina:2013mla} 
  M.~Farina, D.~Pappadopulo and A.~Strumia,
  ``A modified naturalness principle and its experimental tests,''
  JHEP {\bf 1308}, 022 (2013)
  doi:10.1007/JHEP08(2013)022
  [arXiv:1303.7244 [hep-ph]].

\bibitem{Ellis:1986yg} 
  J.~R.~Ellis, K.~Enqvist, D.~V.~Nanopoulos and F.~Zwirner,
  ``Observables in Low-Energy Superstring Models,''
  Mod.\ Phys.\ Lett.\ A {\bf 1}, 57 (1986).

\bibitem{Barbieri:1987fn} 
  R.~Barbieri and G.~F.~Giudice,
  ``Upper Bounds on Supersymmetric Particle Masses,''
  Nucl.\ Phys.\ B {\bf 306}, 63 (1988).

\bibitem{Antoniadis:2014eta} 
  I.~Antoniadis, E.~M.~Babalic and D.~M.~Ghilencea,
  ``Naturalness in low-scale SUSY models and "non-linear" MSSM,''
  Eur.\ Phys.\ J.\ C {\bf 74}, no. 9, 3050 (2014)
  [arXiv:1405.4314 [hep-ph]].

\bibitem{Ciafaloni:1996zh} 
  P.~Ciafaloni and A.~Strumia,
  ``Naturalness upper bounds on gauge mediated soft terms,''
  Nucl.\ Phys.\ B {\bf 494}, 41 (1997)
  [hep-ph/9611204].

\bibitem{de Carlos:1993yy} 
  B.~de Carlos and J.~A.~Casas,
  ``One loop analysis of the electroweak breaking in supersymmetric models and the fine tuning problem,''
  Phys.\ Lett.\ B {\bf 309}, 320 (1993)
  [hep-ph/9303291].

\bibitem{Casas:2003jx} 
  J.~A.~Casas, J.~R.~Espinosa and I.~Hidalgo,
  ``The MSSM fine tuning problem: A Way out,''
  JHEP {\bf 0401}, 008 (2004)
  [hep-ph/0310137].

\bibitem{Casas:2004gh} 
  J.~A.~Casas, J.~R.~Espinosa and I.~Hidalgo,
  ``Implications for new physics from fine-tuning arguments. 1. Application to SUSY and seesaw cases,''
  JHEP {\bf 0411}, 057 (2004)
  doi:10.1088/1126-6708/2004/11/057
  [hep-ph/0410298].

\bibitem{Casas:2005ev} 
  J.~A.~Casas, J.~R.~Espinosa and I.~Hidalgo,
  ``Implications for new physics from fine-tuning arguments. II. Little Higgs models,''
  JHEP {\bf 0503}, 038 (2005)
  doi:10.1088/1126-6708/2005/03/038
  [hep-ph/0502066].

\bibitem{Strumia:1999fr} 
  A.~Strumia,
  ``Naturalness of supersymmetric models,''
  hep-ph/9904247.

\bibitem{Allanach:2006jc} 
  B.~C.~Allanach,
  ``Naturalness priors and fits to the constrained minimal supersymmetric standard model,''
  Phys.\ Lett.\ B {\bf 635}, 123 (2006)
  doi:10.1016/j.physletb.2006.02.052
  [hep-ph/0601089].

\bibitem{Giusti:1998gz} 
  L.~Giusti, A.~Romanino and A.~Strumia,
  ``Natural ranges of supersymmetric signals,''
  Nucl.\ Phys.\ B {\bf 550}, 3 (1999)
  doi:10.1016/S0550-3213(99)00153-4
  [hep-ph/9811386].

\bibitem{Anderson:1994dz} 
  G.~W.~Anderson and D.~J.~Castano,
  ``Measures of fine tuning,''
  Phys.\ Lett.\ B {\bf 347}, 300 (1995)
  [hep-ph/9409419].

\bibitem{Athron:2007ry} 
  P.~Athron and D.~J.~Miller,
  ``A New Measure of Fine Tuning,''
  Phys.\ Rev.\ D {\bf 76}, 075010 (2007)
  doi:10.1103/PhysRevD.76.075010
  [arXiv:0705.2241 [hep-ph]].

\bibitem{Allanach:2007qk} 
  B.~C.~Allanach, K.~Cranmer, C.~G.~Lester and A.~M.~Weber,
  ``Natural priors, CMSSM fits and LHC weather forecasts,''
  JHEP {\bf 0708}, 023 (2007)
  doi:10.1088/1126-6708/2007/08/023
  [arXiv:0705.0487 [hep-ph]].

\bibitem{Cabrera:2008tj} 
  M.~E.~Cabrera, J.~A.~Casas and R.~Ruiz de Austri,
  ``Bayesian approach and Naturalness in MSSM analyses for the LHC,''
  JHEP {\bf 0903}, 075 (2009)
  doi:10.1088/1126-6708/2009/03/075
  [arXiv:0812.0536 [hep-ph]].

\bibitem{Cabrera:2009dm} 
  M.~E.~Cabrera, J.~A.~Casas and R.~Ruiz d Austri,
  ``MSSM Forecast for the LHC,''
  JHEP {\bf 1005}, 043 (2010)
  doi:10.1007/JHEP05(2010)043
  [arXiv:0911.4686 [hep-ph]].

\bibitem{Ghilencea:2012qk} 
  D.~M.~Ghilencea and G.~G.~Ross,
  ``The fine-tuning cost of the likelihood in SUSY models,''
  Nucl.\ Phys.\ B {\bf 868}, 65 (2013)
  doi:10.1016/j.nuclphysb.2012.11.007
  [arXiv:1208.0837 [hep-ph]].

\bibitem{Fichet:2012sn} 
  S.~Fichet,
  ``Quantified naturalness from Bayesian statistics,''
  Phys.\ Rev.\ D {\bf 86}, 125029 (2012)
  doi:10.1103/PhysRevD.86.125029
  [arXiv:1204.4940 [hep-ph]].

\bibitem{Fowlie:2014xha} 
  A.~Fowlie,
  ``CMSSM, naturalness and the "fine-tuning price" of the Very Large Hadron Collider,''
  Phys.\ Rev.\ D {\bf 90}, 015010 (2014)
  doi:10.1103/PhysRevD.90.015010
  [arXiv:1403.3407 [hep-ph]].

\bibitem{Fowlie:2014faa} 
  A.~Fowlie,
  ``Is the CNMSSM more credible than the CMSSM?,''
  Eur.\ Phys.\ J.\ C {\bf 74}, no. 10, 3105 (2014)
  doi:10.1140/epjc/s10052-014-3105-y
  [arXiv:1407.7534 [hep-ph]].

\bibitem{Fowlie:2015uga} 
  A.~Fowlie,
  ``The little-hierarchy problem is a little problem: understanding the difference between the big- and little-hierarchy problems with Bayesian probability,''
  arXiv:1506.03786 [hep-ph].

\bibitem{Athron:2017fxj} 
  P.~Athron, C.~Balazs, B.~Farmer, A.~Fowlie, D.~Harries and D.~Kim,
  ``Bayesian analysis and naturalness of (Next-to-)Minimal Supersymmetric Models,''
  arXiv:1709.07895 [hep-ph].

\bibitem{Clarke:2016jzm} 
  J.~D.~Clarke and P.~Cox,
  ``Naturalness made easy: two-loop naturalness bounds on minimal SM extensions,''
  JHEP {\bf 1702}, 129 (2017)
  doi:10.1007/JHEP02(2017)129
  [arXiv:1607.07446 [hep-ph]].

\bibitem{Fowlie:2016jlx} 
  A.~Fowlie, C.~Balazs, G.~White, L.~Marzola and M.~Raidal,
  ``Naturalness of the relaxion mechanism,''
  JHEP {\bf 1608}, 100 (2016)
  doi:10.1007/JHEP08(2016)100
  [arXiv:1602.03889 [hep-ph]].

\bibitem{AbdusSalam:2015uba} 
  S.~S.~AbdusSalam and L.~Velasco-Sevilla,
  ``Where to look for natural supersymmetry,''
  Phys.\ Rev.\ D {\bf 94}, no. 3, 035026 (2016)
  doi:10.1103/PhysRevD.94.035026
  [arXiv:1506.02499 [hep-ph]].

\bibitem{Kim:2013uxa} 
  D.~Kim, P.~Athron, C.~Balzs, B.~Farmer and E.~Hutchison,
  ``Bayesian naturalness of the CMSSM and CNMSSM,''
  Phys.\ Rev.\ D {\bf 90}, no. 5, 055008 (2014)
  doi:10.1103/PhysRevD.90.055008
  [arXiv:1312.4150 [hep-ph]].

\bibitem{Ghilencea:2012gz} 
  D.~M.~Ghilencea, H.~M.~Lee and M.~Park,
  ``Tuning supersymmetric models at the LHC: A comparative analysis at two-loop level,''
  JHEP {\bf 1207}, 046 (2012)
  doi:10.1007/JHEP07(2012)046
  [arXiv:1203.0569 [hep-ph]].

\bibitem{Dumont:2013wma} 
  B.~Dumont, S.~Fichet and G.~von Gersdorff,
  ``A Bayesian view of the Higgs sector with higher dimensional operators,''
  JHEP {\bf 1307}, 065 (2013)
  doi:10.1007/JHEP07(2013)065
  [arXiv:1304.3369 [hep-ph]].

\bibitem{Ghilencea:2013fka} 
  D.~M.~Ghilencea,
  ``A new approach to Naturalness in SUSY models,''
  PoS Corfu {\bf 2012}, 034 (2013)
  [arXiv:1304.1193 [hep-ph]].

\bibitem{Kvellestad:2015cpa} 
  A.~Kvellestad,
  ``Chasing SUSY Through Parameter Space,''
  Ph.D. thesis, Oslo U. (2015).

\bibitem{Mackay}
D.~Mackay,
``Bayesian Methods for Adaptive Models,''
Ph.D. thesis, California Institute of Technology (1992).

\bibitem{FowlieTalk}
A.~Fowlie,
``Bayesian Approach to Naturalness,''
talk given at conference on ``Fine-tuning, the Multiverse and Life,'' University of Sydney, 2016.
\url{http://www.physics.usyd.edu.au/~luke/2016FTConf/talks/2_1_Fowlie.pdf}

\bibitem{Nesseris:2012cq} 
  S.~Nesseris and J.~Garcia-Bellido,
  JCAP {\bf 1308}, 036 (2013)
  doi:10.1088/1475-7516/2013/08/036
  [arXiv:1210.7652 [astro-ph.CO]].

\bibitem{March}
M.~C.~March, G.~D.~Starkman, R.~Trotta, P.~M.~Vaudrevange,
``Should we doubt the cosmological constant?,''
Mon Not R Astron Soc {\bf 410}, 4 (2011)
doi: 10.1111/j.1365-2966.2010.17614.x\ .

\bibitem{Kunz:2006mc} 
  M.~Kunz, R.~Trotta and D.~Parkinson,
  Phys.\ Rev.\ D {\bf 74}, 023503 (2006)
  doi:10.1103/PhysRevD.74.023503
  [astro-ph/0602378].

\bibitem{Earman}
John Earman,
``Bayes or Bust? A Critical Examination of Bayesian Confirmation Theory,''
The MIT Press, Cambridge, Massachusetts, USA (1992).

\bibitem{Forster}
M.~R.~Forster,
``Bayes and Bust: Simplicity as a Problem for a probabilist's Approach to Confirmation,''
The British Journal for the Philosophy of Science {\bf 46}, 3 (1995).

\bibitem{Oppy}
G. Oppy,
``Bayes not Bust! Why Simplicity is no Problem for Bayesians,''
British Journal for the Philosophy of Science {\bf 58} 4 (2007).

\bibitem{git}
\url{https://github.com/AustinNH/gut-bayesian}

\bibitem{Balazs:2013qva} 
  C.~Balazs, A.~Buckley, D.~Carter, B.~Farmer and M.~White,
  ``Should we still believe in constrained supersymmetry?,''
  Eur.\ Phys.\ J.\ C {\bf 73}, 2563 (2013)
  doi:10.1140/epjc/s10052-013-2563-y
  [arXiv:1205.1568 [hep-ph]].

\bibitem{Jeffreys}
  H. Jeffreys, ``Theory of probability'', 3rd ed., Clarendon Press, Oxford, U.K. (1961)

\bibitem{Olive:2016xmw} 
  C.~Patrignani {\it et al.} [Particle Data Group],
  ``Review of Particle Physics,''
  Chin.\ Phys.\ C {\bf 40}, no. 10, 100001 (2016).
  doi:10.1088/1674-1137/40/10/100001

\bibitem{Kass}
R. E. Kass, L. Wasserman,
``The Selection of Prior Distributions by Formal Rules,''
Journal of the American Statistical Association {\bf 91}, no. 435, 1343 (1996)
doi:10.2307/2291752

\bibitem{Bartlett}
M. S. Bartlett,
``A Comment on D. V. Lindley's Statistical Paradox,''
Biometrika {\bf 44}, no. 3/4, 533 (1957)
doi:10.2307/2332888.

\bibitem{Dimopoulos:1981zb} 
  S.~Dimopoulos and H.~Georgi,
  ``Softly Broken Supersymmetry and SU(5),''
  Nucl.\ Phys.\ B {\bf 193}, 150 (1981).
  doi:10.1016/0550-3213(81)90522-8

\bibitem{Langacker:1990jh} 
  P.~Langacker,
  ``Precision tests of the standard model,''
  In *Boston 1990, Proceedings, Particles, strings and cosmology* 237-269 and Pennsylvania Univ. Philadelphia - UPR-0435T (90,rec.Oct.) 33 p. (015721) (see HIGH ENERGY PHYSICS INDEX 29 (1991) No. 9950)

\bibitem{Ellis:1990wk} 
  J.~R.~Ellis, S.~Kelley and D.~V.~Nanopoulos,
  ``Probing the desert using gauge coupling unification,''
  Phys.\ Lett.\ B {\bf 260}, 131 (1991).
  doi:10.1016/0370-2693(91)90980-5

\bibitem{Amaldi:1991cn} 
  U.~Amaldi, W.~de Boer and H.~Furstenau,
  ``Comparison of grand unified theories with electroweak and strong coupling constants measured at LEP,''
  Phys.\ Lett.\ B {\bf 260}, 447 (1991).
  doi:10.1016/0370-2693(91)91641-8

\bibitem{Langacker:1991an} 
  P.~Langacker and M.~x.~Luo,
  ``Implications of precision electroweak experiments for $M_t$, $\rho_{0}$, $\sin^2\theta_W$ and grand unification,''
  Phys.\ Rev.\ D {\bf 44}, 817 (1991).
  doi:10.1103/PhysRevD.44.817

\bibitem{Giunti:1991ta} 
  C.~Giunti, C.~W.~Kim and U.~W.~Lee,
  ``Running coupling constants and grand unification models,''
  Mod.\ Phys.\ Lett.\ A {\bf 6}, 1745 (1991).
  doi:10.1142/S0217732391001883

\bibitem{Dienes:1996du} 
  K.~R.~Dienes,
  ``String theory and the path to unification: A Review of recent developments,''
  Phys.\ Rept.\  {\bf 287}, 447 (1997)
  doi:10.1016/S0370-1573(97)00009-4
  [hep-th/9602045].

\bibitem{Deen:2016vyh} 
  R.~Deen, B.~A.~Ovrut and A.~Purves,
  ``The minimal SUSY B  L model: simultaneous Wilson lines and string thresholds,''
  JHEP {\bf 1607}, 043 (2016)
  doi:10.1007/JHEP07(2016)043
  [arXiv:1604.08588 [hep-ph]].

\bibitem{Deen:2016zfr} 
  R.~Deen, B.~A.~Ovrut and A.~Purves,
  ``Supersymmetric Sneutrino-Higgs Inflation,''
  Phys.\ Lett.\ B {\bf 762}, 441 (2016)
  doi:10.1016/j.physletb.2016.09.059
  [arXiv:1606.00431 [hep-ph]].

\bibitem{Kaplunovsky:1992vs} 
  V.~S.~Kaplunovsky,
  ``One loop threshold effects in string unification,''
  hep-th/9205070.

\bibitem{Mayr:1993kn} 
  P.~Mayr, H.~P.~Nilles and S.~Stieberger,
  ``String unification and threshold corrections,''
  Phys.\ Lett.\ B {\bf 317}, 53 (1993)
  doi:10.1016/0370-2693(93)91569-9
  [hep-th/9307171].

\bibitem{Dolan:1992nf} 
  L.~Dolan and J.~T.~Liu,
  ``Running gauge couplings and thresholds in the type II superstring,''
  Nucl.\ Phys.\ B {\bf 387}, 86 (1992)
  doi:10.1016/0550-3213(92)90047-F
  [hep-th/9205094].

\bibitem{Kiritsis:1996dn} 
  E.~Kiritsis, C.~Kounnas, P.~M.~Petropoulos and J.~Rizos,
  ``Universality properties of N=2 and N=1 heterotic threshold corrections,''
  Nucl.\ Phys.\ B {\bf 483}, 141 (1997)
  doi:10.1016/S0550-3213(96)00550-0
  [hep-th/9608034].

\bibitem{Klaput:2010dg} 
  M.~A.~Klaput and C.~Paleani,
  ``The computation of one-loop heterotic string threshold corrections for general orbifold models with discrete Wilson lines,''
  arXiv:1001.1480 [hep-th].

\bibitem{deAlwis:2012bm} 
  S.~P.~de Alwis,
  ``Gauge Threshold Corrections and Field Redefinitions,''
  Phys.\ Lett.\ B {\bf 722}, 176 (2013)
  doi:10.1016/j.physletb.2013.04.007
  [arXiv:1211.5460 [hep-th]].

\bibitem{Bailin:2014nna} 
  D.~Bailin and A.~Love,
  ``Reduced modular symmetries of threshold corrections and gauge coupling unification,''
  JHEP {\bf 1504}, 002 (2015)
  doi:10.1007/JHEP04(2015)002
  [arXiv:1412.7327 [hep-th]].

\bibitem{Langacker:1992rq} 
  P.~Langacker and N.~Polonsky,
  ``Uncertainties in coupling constant unification,''
  Phys.\ Rev.\ D {\bf 47}, 4028 (1993)
  doi:10.1103/PhysRevD.47.4028
  [hep-ph/9210235].

\bibitem{Ovrut:2012wg} 
  B.~A.~Ovrut, A.~Purves and S.~Spinner,
  ``Wilson Lines and a Canonical Basis of SU(4) Heterotic Standard Models,''
  JHEP {\bf 1211}, 026 (2012)
  doi:10.1007/JHEP11(2012)026
  [arXiv:1203.1325 [hep-th]].
  
\bibitem{Marshall:2014kea} 
  Z.~Marshall, B.~A.~Ovrut, A.~Purves and S.~Spinner,
 ``Spontaneous R-parity Breaking, Stop LSP Decays and the Neutrino Mass Hierarchy,''
  Phys.\ Lett.\ B {\bf 732}, 325 (2014)
  doi:10.1016/j.physletb.2014.03.052
  [arXiv:1401.7989 [hep-ph]].
  
\bibitem{Marshall:2014cwa} 
  Z.~Marshall, B.~A.~Ovrut, A.~Purves and S.~Spinner,
  ``LSP Squark Decays at the LHC and the Neutrino Mass Hierarchy,''
  Phys.\ Rev.\ D {\bf 90}, no. 1, 015034 (2014)
  doi:10.1103/PhysRevD.90.015034
  [arXiv:1402.5434 [hep-ph]].
  
\bibitem{Ovrut:2014rba} 
  B.~A.~Ovrut, A.~Purves and S.~Spinner,
  ``A statistical analysis of the minimal SUSY BL theory,''
  Mod.\ Phys.\ Lett.\ A {\bf 30}, no. 18, 1550085 (2015)
  doi:10.1142/S0217732315500856
  [arXiv:1412.6103 [hep-ph]].
  
\bibitem{Ovrut:2015uea} 
  B.~A.~Ovrut, A.~Purves and S.~Spinner,
  ``The minimal SUSY $B-L$ model: from the unification scale to the LHC,''
  JHEP {\bf 1506}, 182 (2015)
  doi:10.1007/JHEP06(2015)182
  [arXiv:1503.01473 [hep-ph]].
  
\bibitem{recipes}
William H. Press, Saul A. Teukolsky, William T. Vetterling, and Brian P. Flannery,
``Numerical Recipes 3rd Edition: The Art of Scientific Computing (3 ed.),''
Cambridge University Press, New York, NY, USA (2007)

\bibitem{Manssen:2012vj} 
  M.~Manssen, M.~Weigel and A.~K.~Hartmann,
  ``Random number generators for massively parallel simulations on GPU,''
  Eur.\ Phys.\ J.\ ST {\bf 210}, 53 (2012)
  doi:10.1140/epjst/e2012-01637-8
  [arXiv:1204.6193 [physics.comp-ph]].

\bibitem{philox}
J. K. Salmon, M. A. Moraes, R. O. Dror,	D. E. Shaw
``Parallel Random Numbers: As Easy As 1, 2, 3,''
Proceedings of 2011 International Conference for High Performance Computing, Networking, Storage and Analysis, SC '11 (ACM, New York, NY, USA, 2011)

\end{thebibliography}
\end{document}